\documentclass[graybox]{svmult}

\usepackage{type1cm}
\usepackage{makeidx}
\usepackage{graphicx}
\usepackage{multicol}
\usepackage[bottom]{footmisc}
\usepackage{newtxtext}
\usepackage{newtxmath}
\usepackage{xcolor,colortbl}
\usepackage{pifont}
\usepackage{url}
\usepackage{xspace}
\usepackage{subfig}
\usepackage{wrapfig}
\usepackage{float}

\newcommand{\ie}{\emph{i.e.,}\xspace}
\newcommand{\eg}{\emph{e.g.,}\xspace}
\newcommand{\etc}{etc.\xspace}
\newcommand{\etal}{\emph{et~al.}\xspace} 
\newcommand{\weblink}[1]{\footnote{\url{#1}}}
\newcommand{\ins}[1]{\textcolor{black}{#1}} 

\usepackage{hyperref}
\usepackage{url}
\hypersetup{
    colorlinks=true,
    linkcolor=blue,
    filecolor=magenta,      
    urlcolor=blue,
    citecolor=blue
}

\makeindex

\begin{document}

\title*{Ten Years of Teaching Empirical Software Engineering in the context of Energy-efficient Software}
\author{Ivano Malavolta, Vincenzo Stoico, Patricia Lago}
\institute{Ivano Malavolta, Vincenzo Stoico, Patricia Lago at Vrije Universiteit Amsterdam, The Netherlands, \email{i.malavolta@vu.nl, v.stoico@vu.nl, p.lago@vu.nl}}

\maketitle

\abstract{In this chapter we share our experience in running ten editions of the Green Lab course at the Vrije Universiteit Amsterdam, the Netherlands. The course is given in the Software Engineering and Green IT track of the Computer Science Master program of the VU. The course takes place every year over a 2-month period and teaches Computer Science students the fundamentals of Empirical Software Engineering in the context of energy-efficient software. \\
The peculiarity of the course is its \textit{research orientation}: at the beginning of the course the instructor presents a catalog of scientifically relevant goals, and each team of students signs up for one of them and works together for 2 months on their own experiment for achieving the goal. Each team goes over the classic steps of an empirical study, starting from a precise formulation of the goal and research questions to context definition, selection of experimental subjects and objects, definition of experimental variables, experiment execution, data analysis, and reporting. \\
Over the years, the course became well-known within the Software Engineering community since it led to several scientific studies that have been published at various scientific conferences and journals. Also, students execute their experiments using \textit{open-source tools}, which are developed and maintained by researchers and other students within the program, thus creating a virtuous community of learners where students exchange ideas, help each other, and learn how to collaboratively contribute to open-source projects in a safe environment.}

\section{Introduction}\label{sec:intro}
 
Developing energy-efficient software is not an option anymore, it is a matter of survival and moral duty.
The Track Clean Energy Progress (TCEP)\footnote{The Track Clean Energy Progress is a report published by the International Energy Agency (IEA) that evaluates the latest progress regarding the transition to clean energy sources}
report of 2023 indicates that data centers and data transmission networks contribute, respectively, to the 1-1.5\% of the global energy demand \cite{tcep_iea}. This corresponds to approximately 240-340 TWh for data centers alone, more than double the amount needed to power the Netherlands in 2020 (i.e., 115.88 TWh). Despite evidence that technological improvements helped mitigate their energy demand in the past years, the request for data center resources and network traffic is increasing. Especially with the expansion and large-scale application of AI-based solutions like generative AI (\eg to create new contents like text, images, and videos)~\cite{tcep_iea,generative_ai}, there is growing evidence that the electricity consumption (and likely the carbon footprint) of software will escalate in the next years~\cite{de2023growing,verdecchia2023systematic}. 

With such high stakes, it is fundamental that practitioners follow an \textbf{evidence-based approach} and make informed decisions about the energy efficiency and carbon footprint of their software~\cite{manotas2016empirical,pinto2017energy}. 
In this context, thanks to its roots in insight- and evidence-oriented nature, the importance of Empirical Software Engineering~\cite{wohlin2012experimentation} is already prominent among both researchers and practitioners. 
However, applying empirical methods to assess the energy aspects of software is a difficult endeavour; precisely measuring the energy efficiency of software in a reliable manner requires strong technical skills and it has proven to be very challenging, time-consuming, and technologically fragmented~\cite{IST_2022_2,georgiou2019software}.  
As such, strong competencies in empirical methods applied to energy-efficient software are in high demand~\cite{saraiva2021bringing}. 

This chapter is about our experience in running 10 editions of the \textbf{Green Lab} course at the Vrije Universiteit (VU) Amsterdam, the Netherlands. The course is given in the Software Engineering and Green IT (SEG) track of the Computer Science Master program of the VU. The course takes place every year over a 2-month period and teaches Computer Science students the fundamentals of Empirical Software Engineering (ESE) in the context of energy-efficient software.

The main peculiarity of the course is its \textbf{research orientation}: at the beginning of the course the instructor presents a catalog of scientifically relevant problems, and each team of students signs up for one of them and works together for 2 months on their empirical study to solve the assigned problem. Each team goes over the classic steps of an empirical study (a la Wohlin \etal~\cite{wohlin2012experimentation}), starting from the formulation of the goal and research questions to context definition, selection of experimental subjects and objects, definition of experimental variables, experiment execution, data analysis, and reporting.
Over the years, the course became well-known within the international Software Engineering research community since it led to several scientific studies published at various scientific conferences and journals, such as EASE, MOBILESoft, IST journal, ICT4S, \etc, even winning best and distinguished paper awards. 

The course has a \textbf{hands-on nature}. Student projects (i) use real-world software products as experimental subjects (\eg by mining the frontend of publicly-available websites, mining mobile apps, or bringing up widely-used open-source platforms) and  (ii) involve the execution and measurement of the experimental subjects on dedicated hardware (\eg smartphones, servers, laptops, IoT devices). The hands-on nature of the course exposes students to recurrent concerns (and their corresponding trade-offs) that software engineering researchers have when designing and conducting controlled experiments on the energy-efficiency of software. Examples of such concerns include: the selection of representative objects/subjects, reliability and precision of the energy measurement tools, feasibility of the experiment (and corresponding decisions in terms of experiment design), correctness of the statistical analysis, representativeness of the obtained results.

Students execute their experiments using \textbf{open-source tools} (\eg Android Runner~\cite{A_Mobile_2020}), which are developed and maintained by the instructors of the course and other students within the program, thus creating a virtuous community of learners where students exchange ideas, help each other, and learn how to collaboratively contribute to open-source projects in a safe environment.

The \textbf{intended target audience} of this chapter are instructors willing to include research- and empirical-oriented activities into Computer Science and Software Engineering programs.
We designed the Green Lab course as a \textit{technology-independent course}, \ie 
its learning objectives, structure, assessment procedure, and lectures are exclusively about energy efficiency and ESE research
methods, and not bound to any specific technology or application domain.
Instructors can access the study guides of the course, its slide sets, links to replication packages of experiments, and links to open-source tools in the \textbf{supplementary online material} of this chapter~\cite{replication_package}.

The remainder of this chapter is organized as follows.
Section~\ref{sec:educational_context} presents the educational context in which the Green Lab course is embedded. Section~\ref{sec:course_design} presents the design of the course, its learning objectives, and main educational components.
Section~\ref{sec:equipment} describes the equipment and measurement infrastructure supporting the experiments developed in the context of the Green Lab course. 
Section~\ref{sec:tools} presents the open-source tools developed by the Green Lab community.
Representative examples of scientific studies emerging from the Green Lab are presented in Section~\ref{sec:success_stories}, while we elaborate on lessons learned for instructors in Section~\ref{sec:lessons_learned}.
We wrap up the chapter and provide our planned future improvements of the course in Section~\ref{sec:conclusions}. 
\section{Educational Context}\label{sec:educational_context}

As mentioned in Section~\ref{sec:intro}, the Green Lab course is part of the SEG track which provides one of the possible specializations within the 2-year Computer Science Master program at VU Amsterdam. 
At the time of creating the track~\cite{Lago2014-SEGTrack}, the relation between software engineering and energy was barely understood. Green IT, instead, was a term emerging yet resonating among people. Hence, we decided to name the track after this term. However, we used it in its broad definition (so as to include software, too), \ie ``\textit{the study and practice of designing, manufacturing, using, and disposing of computers, servers, and associated subsystems efficiently and effectively with minimal or no impact on the environment}''~\cite{Murugesan2008-rg}.

The SEG track entails five so-called \textit{core} courses. Each course counts 6 credits according to the European Credit and Transfer Accumulation System (ECTS), hence counting a total of 30 ECTS. To integrate energy awareness in the track, we have adopted a mix of the \textit{distributed approach} and the \textit{centralized approach} identified by Mann \etal \cite{Mann2010-qb}. Accordingly, we both revisited pre-existing courses across the whole track (\textit{distributed approach}) and created a few dedicated specific courses \textit{focusing on sustainability whilst also having sustainability issues addressed across the curriculum} (blended approach)~\cite{Mann2010-qb}. The Green Lab belongs to the latter, by teaching competencies that \textit{blend} empirical software engineering in the specific context of energy-efficient software).

Further, we adopted a distributed approach for the other core courses Service Oriented Design, Digital Architecture, and Fundamentals of Adaptive Software (by applying software engineering competencies to practical projects and assignments that include Green IT aspects); and a centralized approach for core course Software Testing (by teaching traditional software engineering competencies that can possibly be used to test software for \eg detecting energy hotspots~\cite{Procaccianti2015-greenlab}). 

Timewise, the academic year at VU Amsterdam is organized into 2 semesters, each including 3 periods of 2-2-1 month of 8-8-4 weeks, respectively, where the last week is dedicated to exams. From a student perspective, an 8-week period should include two courses that require 50\% of the student's time, each, while a 4-week period is dedicated to a single course full-time. For instance, the Green Lab course is taught in an 8-week period, in parallel with another course.

Finally, according to the academic level of their learning objectives, courses are categorized as at specialized (400), research-oriented (500), or highly specialized (600) level. For instance, from the academic year 2024/2025 the Green Lab course will be classified as a 500-level course thanks to the consolidation of energy-efficiency software engineering and measurement as fundamental research competencies. 
\section{Course Design}\label{sec:course_design}
This section presents the main educational components of the current edition of the course, including the students' cohort, learning objectives, and teaching team (Section~\ref{sec:audience}), course contents and structure (Section~\ref{sec:contents}), and assessment method (Section~\ref{sec:project}). 

\subsection{Students' cohort, Learning Objectives, and Teaching Team}\label{sec:audience}
In terms of \textbf{students' cohort}, this course has been designed with Computer Science Master students in mind. 
Given the technical nature of the course, the recommended background knowledge of students attending the course includes: (i) basic statistical analysis techniques (\ie descriptive statistics) and most common tests and (ii) basic programming/scripting skills. Both requirements are not formally enforced when students enroll in the course, but the lecturer informs students about them in the first lecture of the course, so to set the expectations and to allow students to prepare for the rest of the course. Students are not expected to be knowledgeable of ESE research methods, which are part of the learning objectives of the course. 

In the last edition of the course (2023/2024 academic year), the course was attended by 91 active students. The course attracts a variety of students coming from several programs and Master tracks. Specifically, within the Computer Science Master program of VU Amsterdam, students' provenance is composed as follows: 43 students of the Software Engineering and Green IT Master track described in Section~\ref{sec:educational_context} and 18 students of other tracks within the same Computer Science Master program (14 in the Big Data Engineering track, 2 in the Foundational Computing and Concurrency track, and 2 in the Internet and Web Technology track);
a total of 24 students come from international tracks of the  Computer Science Master program, of which 20 come from the Software Engineering for the Green Deal (SE4GD) track and 4 come from the Global Software Engineering European Master (GSEEM) track  
and 1 in the Parallel Computing Systems track); finally, 5 students come from other Master programs, including (i) Parallel Computing Systems at VU Amsterdam, (ii) Computer Systems Security at VU Amsterdam, Human Computing Interaction at Utrecht University, and Computer Science at the University of Zurich. 

\begin{wrapfigure}[14]{r}{0.4\textwidth}
  \begin{center}
    \vspace{-1cm}
    \includegraphics[width=0.4\textwidth]{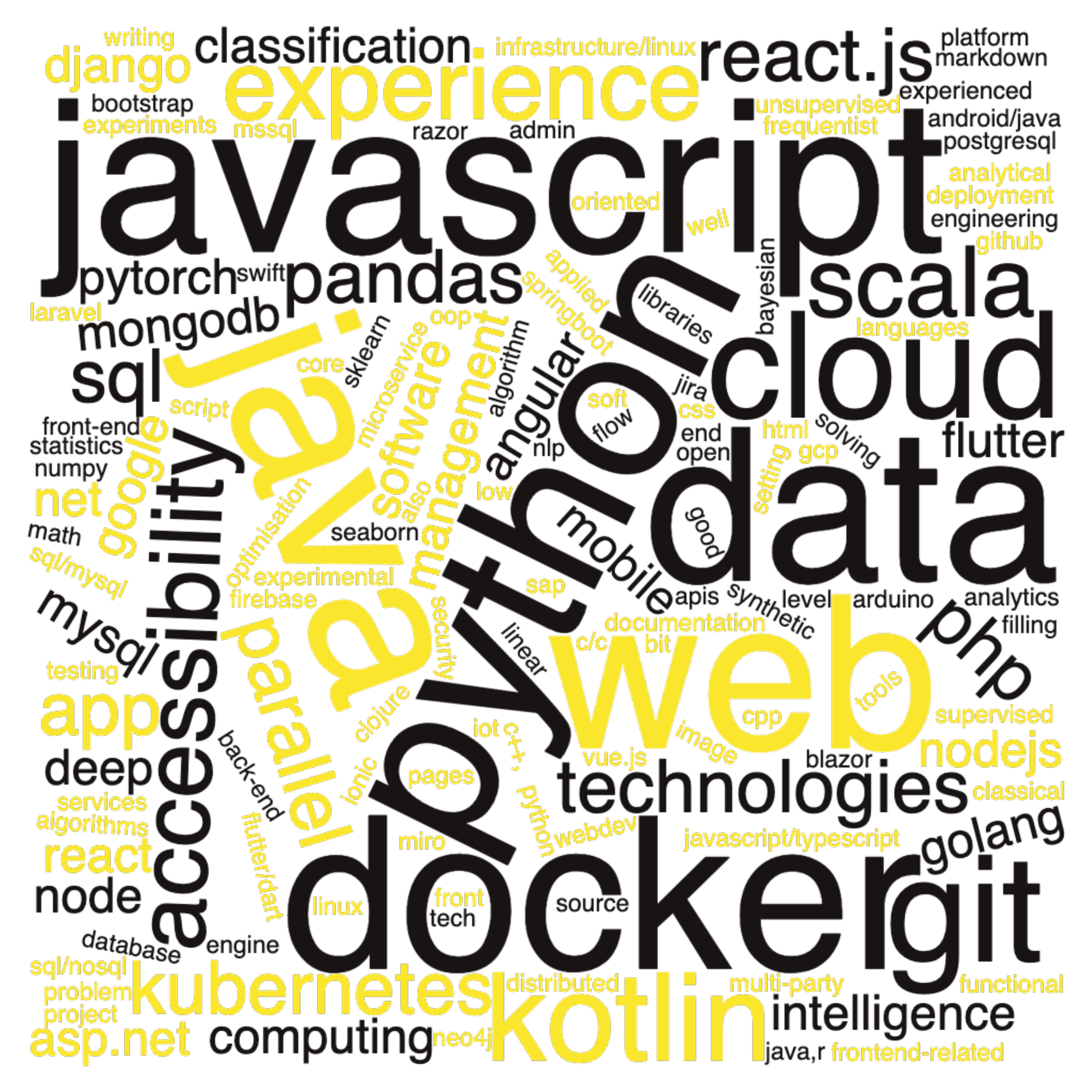}
    \vspace{-1cm}
  \end{center}
  \caption{Technical skills of Green Lab students in 2023/2024}
  \label{fig:tech_skills}
\end{wrapfigure}
Over the years, students' ages, genders, nationalities, scientific backgrounds, and technical skills became more and more heterogeneous. For the sake of students' privacy we do not report students' personal demographics. Figure~\ref{fig:tech_skills} shows the most recurrent technical skills of the students enrolled in the last edition of the course. The heterogeneity of students' technical skills and backgrounds has been taken care when designing the course and mitigated in the assessment phase of the course by allowing students to work in their projects with the technologies they are more familiar with (see Section~\ref{sec:project}).    

Upon completion of this course, the \textbf{learning objectives} in Table~\ref{tab:learning_objectives} are achieved. In the table, we report also the levels of learning covered by each learning objective according to the revised version of Bloom’s taxonomy of learning~\cite{bloom2020taxonomy,anderson2001taxonomy}. 

\begin{table}
    \caption{Learning objectives of the Green Lab course}
    \label{tab:learning_objectives}
    \centering
    \begin{tabular}{|c|p{6cm}|p{4cm}|}
        \hline
        \textbf{ID} & \textbf{Learning objective}  & \textbf{Bloom's levels of learning} \\
        \hline
        LO1 & To learn the principles of empirical experimentation in the field of Software Engineering & Remembering -- Understanding \\ \hline
        LO2 & To measure and assess the impact of software on energy consumption and environmental footprint. & Understanding -- Applying \\ \hline
        LO3 & To become familiar with and critically reflect on the research results, trends, and challenges in the field of green software engineering. & Understanding -- Analyzing  \\ \hline
        LO4 & To design and conduct successful experiments on the energy efficiency of software. & Applying -- Analyzing -- Evaluating -- Creating \\ \hline
        LO5 & To synthesize, correctly interpret, and put into context the results of an experiment on the energy efficiency of software. & Evaluating \\ \hline

    \end{tabular}
\end{table}

The \textbf{teaching team} is composed of two lecturers and three teaching assistants. The responsibilities of the five members of the teaching team are:
\begin{itemize}
    \item Lecturer 1: course coordination, giving all lectures on ESE methods (see Section~\ref{sec:contents} for the details), acting as the main contact point for students and colleagues, assessment of 35\% of students' projects.
    \item Lecturer 2: giving the lecture on green software (see Section~\ref{sec:contents} for the details), supporting students experiencing technical issues with respect to servers and IoT devices, advising students in choosing and configuring measurement tools, assessment of 35\% of students' projects. 
    \item Teaching assistant 1: giving the lab on the Green Lab environment, tools, and measurement tools (lab 1 -- see Section~\ref{sec:contents}), supporting students experiencing technical issues with respect to mobile devices and apps, assessment of 10\% of students' projects.
    \item Teaching assistant 2: giving labs on data analysis with R\weblink{https://www.r-project.org} (labs 2, 3, and 4 -- see Section~\ref{sec:contents}), assessment of 10\% of students' projects.
    \item Teaching assistant 3: supporting students experiencing general technical issues, assessment of 10\% of students' projects.
\end{itemize}

In the 2023/2024 edition of the course, lecturer 1 is an Associate Professor in Software Engineering, with experience on ESE applied to energy-efficient software, lecturer 2 is a postdoc researcher specialized in software performance engineering and energy-efficient software; lecturer 2 is also responsible for the daily activities within the Green Lab servers (see Section~\ref{sec:equipment}). The three teaching assistants are Master students; teaching assistant 1 is also the main developer of the open-source tools used in the Green Lab course (see Section~\ref{sec:tools}). 
The structure of the teaching team emerged organically over the years and it is influenced by (i) the growing number of students in the course\footnote{The first edition of the course had 5 students, five years ago it had 21 students, and in its last edition (2023/2024) it had 91 students.}, (ii) available resources, (iii) the specific research interests of involved lecturers (and teaching assistants, to a lesser extent).    

\subsection{Course Contents and Structure}\label{sec:contents}

\begin{figure}
  \begin{center}
    \includegraphics[width=1\textwidth]{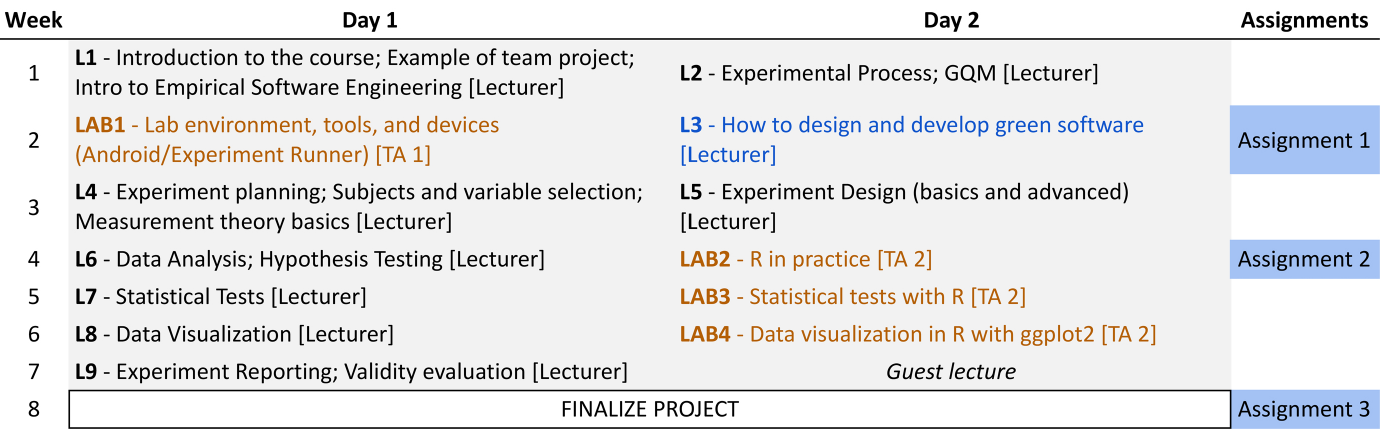}
  \end{center}
  \caption{Schedule of the course in 2023/2024}
  \label{fig:schedule}
\end{figure}

The course is designed to expose students to the fundamentals of Empirical Software Engineering (learning objective LO1) in the context of energy-efficient software (learning objectives LO2, LO3, LO4, and LO5). As such, the course covers first the basic principles of ESE (\eg focus on applicability, quantitative vs qualitative research methods); then, given the importance of measurement-based studies in the context of energy-efficient software, it delves into the specifics of the \textit{controlled experiment} method with software subjects and objects (as opposed to experiments with human subjects~\cite{ko2015practical}). The overall organization of the course is presented in Figure~\ref{fig:schedule}.

As anticipated in Section~\ref{sec:educational_context}, the duration of the course is 8 weeks, with lectures and labs being given in the first 7 weeks. In each of the first 7 weeks, there are two 2-hours educational components in which the teaching team gives either a lecture or a lab. The majority of the lectures (depicted in black in Figure~\ref{fig:schedule}) are about ESE research methods; those lectures are designed so that (i) their main contents and principles are generic and (ii) the lecturer systematically refers to examples, scientific results, cases, and success stories related to energy-efficient software. This makes the course \textit{future-proof} for both students and lecturers: (i) students are able to apply the learned ESE research methods also in contexts different from energy-efficient software (\eg in their final thesis or during their professional career) and (ii) every year, lecturers update only the energy-specific material (\eg scientific studies, cases) without needing to completely restructure the course.
The course includes also a special lecture about how to design and develop green software (L3 -- depicted in blue in Figure~\ref{fig:schedule}); this lecture emerged from the students' feedback we received in the previous years of the course, where students were signaling the fact that the course was ``too meta'' on how to measure the energy-efficiency of software, instead of providing concrete advice on how to make software more energy efficient. During lab sessions (depicted in orange in Figure~\ref{fig:schedule}), students are assisted for technical operation of the lab equipment as regards measurement and tools (see Sections \ref{sec:equipment} and \ref{sec:tools} for the details). In the labs, students also receive the required training for data analysis and visualization using R and RStudio\weblink{https://posit.co/download/rstudio-desktop}. In order to provide students with the wider perspective on ESE methods and/or energy-efficient software, the last educational component of the course is a guest lecture. The guest lectures we had in the years are: \textit{Leveraging Implementation Diversity to Improve Energy Efficiency} (2023, Fernando Castor -- University of Twente), \textit{Sustainability and Cloud} (2022, Thiago de Faria, Amazon Web Services), \textit{Energy-efficient robotic software} (2021, Michel Albonico, Federal University of Technology Paraná).

Throughout the course, students work in teams to perform experiments on energy-efficient software. Students carry out all the phases of a controlled experiment, from experiment design to execution, data analysis, and reporting. Students are provided with examples of experiments coming from state-of-the-art literature, but they will have to choose by themselves the experimental subjects and hypotheses to test. Further details about the project are given in Section~\ref{sec:project}.

At the end of every lecture, the lecturer provides a series of (mandatory and optional) readings. The \textbf{syllabus} of the course is defined as follows:
\begin{itemize}
    \item Textbook: The book ``Experimentation in Software Engineering'' by Wohlin \etal~\cite{wohlin2012experimentation} is the main textbook of the course and the majority of the contents of given lectures follow the same structure of the book. 
    \item Complementary mandatory readings: the textbook is complemented by (i) chapter 6 and 8 of the book ``Guide to advanced Empirical Software Engineering''~\cite{shull2007guide} about statistical methods, measurement theory, and experiment reporting and (ii) chapters 5 and 16 of the book ``Basics of software engineering experimentation''~\cite{juristo2013basics} about designing and reporting empirical studies. 
    \item Support material: additional sources are provided to the students as optional material to be used as a manual/documentation in case students need to go deep on specific aspects of their projects. Specifically:
    \begin{itemize}
        \item Material on ESE research methods: a selection of scientific studies and guidelines for conducting empirical studies. For example, such studies are about statistical analysis in ESE~\cite{dybaa2006systematic,de2019evolution}, data analysis in ESE~\cite{vegas}, experiment design in ESE~\cite{pfleeger1995experimental}.  
        \item Material on energy efficiency: a selection of studies providing either the bigger picture about energy-efficient software or methodological guidelines. For example, such studies are about Green IT and green software in general~\cite{verdecchia2021green},  practitioners' perspectives on green software~\cite{manotas2016empirical,pinto2017energy}, and guidelines for energy measurement and its related issues~\cite{ardito2019methodological,pramanik2019power}. 
        \item Material on statistics and data visualization: a selection of studies and guidelines about statistics and data visualization in general, \eg \cite{sheskin2003handbook} and \cite{seltman2018experimental}. 
    \end{itemize}
    \item Articles on well-conducted experiments (\eg \cite{santos2018does,linares2013api,chowdhury2018exploratory}): a selection of well-designed and reported empirical studies; the lecturer use such studies in class as success stories and as triggers for critical reflection. This part of the syllabus also includes scientific studies emerging from previous editions of the Green Lab (see Section~\ref{sec:success_stories}); those studies help in setting the standard for the expected quality of students' projects.  
    \item Programming books: books about R~\cite{wickham2023r}, Python~\cite{lutz2013learning}, and Bash programming~\weblink{https://www.systemcodegeeks.com/minibook/linux-bash-programming-cookbook}; these are the programming languages that are very likely used across all projects. Students are invited to use those books to refresh their technical skills and solving specific issues they might have during the project.
    \item Videos: a selection of YouTube videos recorded by researchers for scientific conferences like ICSE and MSR\footnote{Using videos also gives to students a ``human'' touch to the scientific studies they read during the course, remembering them that also researchers are human beings and sometimes they are just a few years older than them}.  
\end{itemize}

The course counts as 6 ECTS credits. Since each ECTS credit is equivalent to 28 hours of study, the total \textbf{expected effort of each student} is 168 hours. To set students' expectations right, in the first lecture of the course we advise the following breakdown of activities: (i) in each of the first 7 weeks the student invests 4 hours for the two educational components, 8 hours of study, and 8 hours of work on the project, (ii) the remaining 28 hours are used for finalizing the project in the final week of the course.   

\subsection{Assessment}\label{sec:project}

Teams of 5 students conduct a concrete research project throughout the whole course. Each part of the research project is discussed during the lectures, started during the labs, and completed as homework, so to keep students on track within the course schedule.

The goal of the research project is to plan, design, conduct, and report a scientific experiment in the context of the energy efficiency of various types of software (\eg mobile apps, software libraries, microservices). Each team works on a specific topic in the context of energy-efficient software. In this way, students put into practice the skills and techniques that they have learned during the lectures/labs and develop their practical insights by applying them to real software. In the first lecture of the course, the lecturer provides a list of possible topics, then each team indicates (i) its members, (ii) the technical skills and educational background of each member, and (iii) a set of scores indicating the preferences about each proposed topic. Examples of potentially assignable topics are (i) What is the energy efficiency of AI-generated algorithms with respect to human-created algorithms? (ii) How do configuration settings of the Zipkin monitoring tool impact the energy consumption of a Docker-based system? (iii) How does WebGPU compare against WebGL in terms of the energy efficiency of web apps?
The topics are assigned to teams based on the preferences, technical skills, and knowledge indicated by each team after the first lecture of the course. Then, each team is responsible for independently carrying on the experiment on the assigned topic. When possible, the lecturer provides relevant datasets, scripts, and other material to the team in order to smooth the execution of their experiment. \ins{Such information is typically provided during the first lecture of the course (contextually to the description of the topics for students' projects), during the first lab, during the in-person discussions with students before or after lectures/labs, or contextually to the grading (\ie as part of the feedback given to students on their assignments); moreover, in case a new relevant dataset, tool, or scientific study is published, lecturers make an announcement on the web platform of the course about it, so that students can evaluate whether the new  material can be used in their project.}


As shown in Figure~\ref{fig:schedule}, each team project is composed of 3 assignments. Each assignment deals with a specific increment of the same research project. The final assignment is a fully-fledged scientific study, including all phases of a classical ESE study reporting on a measurement-based controlled experiment.
The sequence of assignments is designed in such a way that their difficulty and required effort grow together with students' deeper understanding of ESE principles, the nuances of their research project, and their experience in terms of energy measurement. We believe that such incremental design of the research project is important to create a safe environment for students to explore and study at the beginning of the course, and then let the project grow together with them during the course.

All assignments include a \textbf{written report} describing all the information related to a specific phase of the project. A complete template for Green Lab reports is publicly available on Overleaf\weblink{https://www.overleaf.com/latex/templates/green-lab-report-template/cpchhrgcrnrr}. Written reports contain also a link to a \textit{time log}, where students record their time; the time log is used only in case of disputes among team members or in case of strong suspicions of fraud. 

Below we describe the main components of each of the 3 assignments.

\begin{itemize}
    \item \textbf{Assignment 1 -- Context and experiment definition} (20\% of the final grade). This part is composed of three sub-components:
    \begin{itemize}
        \item \textit{Technical/societal context}: students describe (i) the domain (\eg mobile apps and their market) and the technologies relevant to the experiment, (ii) the main motivation behind the experiment, including examples, apps/tools screenshots, snippets of source code, \etc, (iii) a brief overview of the direction of the experiment, and (iii) what the target audience (\eg software developers) will learn from the results of the experiment.
        \item \textit{Scientific context}: students position their research by elaborating on 5 to 8 related studies that are similar in terms of goal or methodology, with a clear characterization of the novelty of their own experiment with respect to them.
        \item \textit{Experiment definition}: students use the Goal-Question-Metric (GQM) approach \cite{gqm,gqm2} to formulate the goal, research questions, and high-level metrics (\eg power, execution time, energy, CPU usage) of their experiment. Using the GQM approach helps students (and instructors) get a complete and concise overview of the main direction of the experiment. 
    \end{itemize}
    \item \textbf{Assignment 2 -- Experiment planning and measurement infrastructure setup} (30\% of the final grade). This part is composed of two sub-components:
    \begin{itemize}
        \item \textit{Experiment planning}: students make a detailed plan about the experiment in terms of (i) subjects' selection/mining, (ii) experimental variables (\ie dependent and independent variables), (iii) formal definition of experimental hypotheses, (iv) experiment design (\ie main, co-, fixed, and blocking factors and their organization), (v) data analysis procedure. In this phase, students also provide an estimation of the total duration of the experiment considering the total number of subjects, trials, runs, and estimated execution time for each run; over the years, this proved to be a valuable tool for facilitating students' reasoning on the feasibility of their experiment and take countermeasures, \eg by reducing the number of subjects or moving from a full-factorial design to an incomplete one. In general, we advise students to design their experiment in such a way that its execution takes less than 40 hours. 
        \item Measurement infrastructure setup: students present the technical aspects for executing the experiment, which tools they are going to use, which devices/servers, and the software/hardware infrastructure they are setting up for the experiment.
    \end{itemize}
    \item \textbf{Assignment 3 -- Experiment execution, data analysis, and reflection} (50\% of the final grade). This part is composed of five sub-components: 
    \begin{itemize}
        \item \textit{Experiment execution}: students execute the experiment and collect measures as planned in the second assignment (after having addressed the instructor's feedback). The output of this phase is the raw data collected from the executed experimental runs.
        \item \textit{Data analysis and results reporting}: students check the correctness of the collected measures, compute descriptive statistics, perform the statistical analysis for hypotheses testing (with effect size estimation), and synthesize the obtained results in their report providing suitable plots and tables to illustrate the main points of attention.
        \item \textit{Threats to validity}: students reflect and report the threats to the validity of the experiment according to the Cook and Campbell classification~\cite{Cook_Campbell_1979}.
        \item \textit{Reflection}: students report the main implications and interpretations of the obtained results (possibly grouped by research question). In this phase, since they spent weeks on the technical aspects of the experiment, students tend to have a narrow perspective and focus very much on the low-level, technical aspects of the experiment. So, here students are asked to address directly the target audience of their experiment (\ie who will benefit from the results of the experiment), as described in the previous assignments; this helps students in getting a wider perspective about their experiment and the general usefulness of their results.  
        \item \textit{Report finalization}: students wrap up the project by doing a final complete pass on the report and writing the abstract and conclusion section.
    \end{itemize}
\end{itemize} 

Assignment 3 includes also a link to a GitHub repository containing the complete \textbf{replication package} of the experiment. The replication package must contain: (i) source code of the scripts developed for running the experiment, (ii) source code of any software developed for building used dataset, (iii) raw data resulting from the execution of the experiment, (iv) R scripts for data analysis, and (v) any other relevant material for replicating the experiment. A template for replication packages is provided at the beginning of the course in the form of a GitHub repository\weblink{https://github.com/S2-group/template-replication-package}. Assignment 3 includes also a link to a \textbf{video} where the team presents the main aspects of the experiment in a complete manner (from the motivation and context, design, execution, to the results, discussion, \etc). Presentations are prepared as groupwork where each student presents individually a part of the experiment. Presentations are assessed as either \texttt{pass} or \texttt{fail}.

When grading each assignment, the teaching team provides (i) a quantitative assessment of each part of the assignment according to a shared rubric (available in the supplementary material of this chapter) and, more importantly, (ii) detailed feedback on how to improve the experiment. Intermediate assignments are evaluated and are part of the final assessment of the whole team project (assignment 3). Assignment 3 is a coherent integration of the previous assignments. When working on assignments 2 and 3, student teams are requested to address the feedback provided by the teaching team in the preceding assignment. Over the years, feedback on intermediate assignments proved extremely useful since it allowed the teaching team to (i) ensure the feasibility of the project (in some cases, students tended towards over-promising), (ii) ensure that the difficulty of assigned projects is similar among all projects within the cohort of students, and (iii) steer projects towards scientifically challenging and novel research directions.


\section{Equipment and Measurement Infrastructure} 
\label{sec:equipment}
\begin{figure}
    \centering
    \subfloat[\centering Green Lab physical servers\label{fig:cluster}]{{\includegraphics[width=4.5cm]{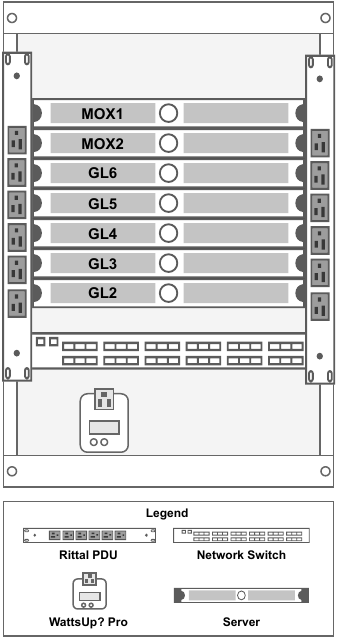}}}%
    \qquad
    \subfloat[\centering Green Lab -- HPC configuration \label{fig:hpc}]{{\includegraphics[width=6.6cm]{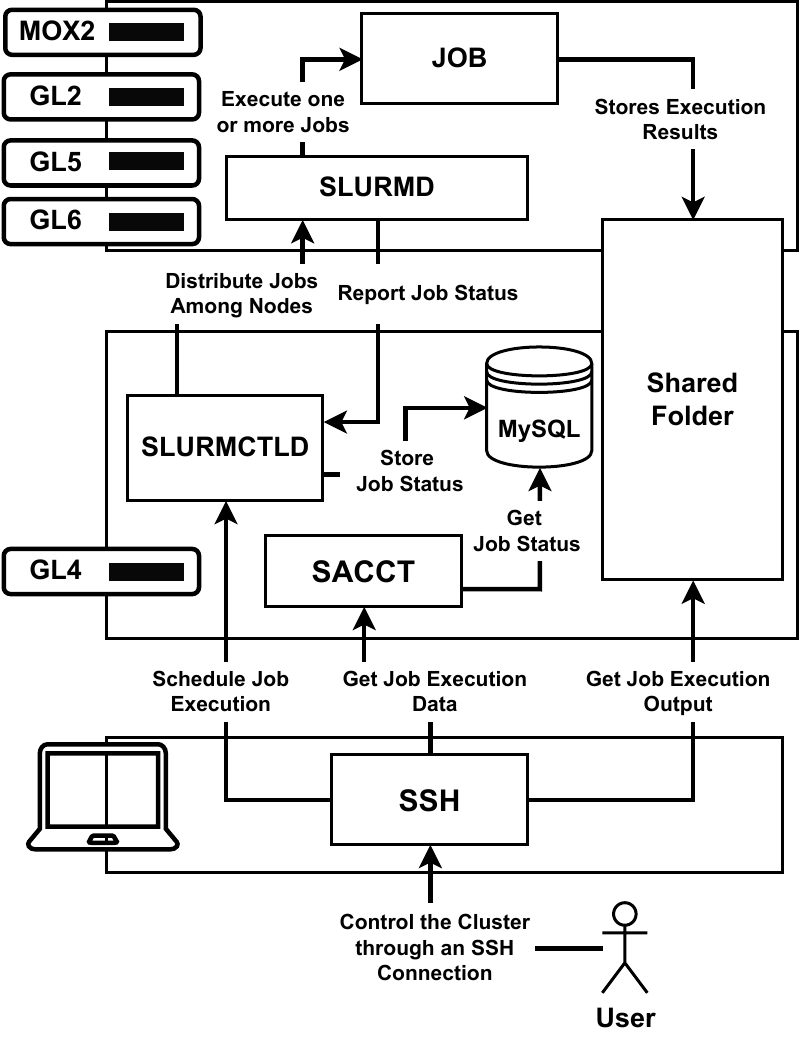}}}%
    \caption{The Green Lab physical infrastructure}%
    \label{fig:greenlab}%
\end{figure}

Green Lab empowers researchers and students to explore diverse software-related domains. So far, the Green Lab has been used for experiments on Distributed, Mobile, Robotics, Artificial Intelligence (AI), Virtual Reality (VR) and Embedded software applications. 
The Green Lab includes a \textit{cluster} of 7 servers with different specifications. Table \ref{tab:cluster} presents an overview of the hardware specifications of each server, along with the operating system currently in use, while Figure \ref{fig:cluster} represents the cluster rack. MOX1, MOX2, GL5, and GL6 are the most powerful servers in the cluster. The cluster supports different ways of executing the software and managing the resources of the servers. Indeed, it is configured to run virtual machines and containers, deploy an application on a single node, as well as distribute and parallelize software across the cluster. The cluster enables users to isolate software applications on a testbed and control the factors influencing their energy consumption and performance. For example, users might be interested in generating different synthetic workloads and injecting them into the application running in our cluster. As a result, it is possible to discover potential energy/performance hotspots and understand how the application reacts under stress.
\begin{table}
\centering
\footnotesize{
\caption{Green Lab Servers Specification}
\label{tab:cluster}
\begin{tabular}{|p{1.5cm}|p{1.5cm}|p{1.5cm}|p{4cm}|p{2.5cm}|}
\hline \textbf{ID}        & \textbf{HD} & \textbf{RAM} & \textbf{CPU} (Intel Xeon)       & \textbf{Operating System}  \\ \hline
MOX1               & 36Tb        & 196Gb        & Silver 4112@2.60GHz  & Debian 9     \\ \hline
MOX2               & 36Tb        & 384Gb        & Silver 4208@2.10GHz  & Debian 11    \\ \hline
GL6                & 36Tb        & 384Gb        & Silver 4208@2.10GHz  & Ubuntu 20.04 \\ \hline
GL5                & 36Tb        & 384Gb        & Silver 4208@2.10GHz  & Ubuntu 22.04 \\ \hline
GL4                & 1Tb         & 16Gb         & E5335@2.00GHz        & Ubuntu 22.04 \\ \hline
GL2                & 1Tb         & 32Gb         & E3-1231@3.40GHz      & Ubuntu 22.04 \\ \hline
GL3                & 126Gb       & 8Gb          & E5345@2.33GHz        & Ubuntu 22.04 \\ \hline
\end{tabular}
}
\end{table}


MOX1 runs Proxmox \cite{goldman2016learning}, an open-source server virtualization software, and, therefore, is dedicated to \textit{virtualization}. Proxmox enables students and researchers to create virtual machines on-demand for testing software systems, executing data analysis, and monitoring other servers. Proxmox supports two virtualization technologies: Kernel-based Virtual Machines and Linux Containers. Moreover, it allows dynamic scaling of resources assigned to each virtual machine, enabling the deployment of different-size workloads at run-time. Proxmox offers a web-based interface for creating, removing, and allocating resources to virtual machines remotely. 
The cluster also supports \textit{containerization}. Indeed, container-based applications, such as SockShop \cite{2023_sock} and Train Ticket Booking System (TTBS) \cite{2023_ttbs}, can be deployed either on a single machine and across the servers. 

The cluster is configured to offer all the computational resources of the servers, and, thus, supports \textit{High-Performance Computing} (HPC) applications that need to execute compute-intensive workloads, such as simulation, machine learning models, and data analytics. To do this, we use SLURM \cite{yoo2003}, an open-source workload manager that handles the distribution and parallelization of computation across the nodes of the cluster. SLURM allows the users to create and distribute computation and monitor the energy consumption and the performance of the jobs and of the nodes within the cluster. Figure \ref{fig:hpc} describes a typical execution of a job on our cluster. Figure \ref{fig:hpc} shows the current configuration of our HPC cluster, which includes a head node, \ie GL4, and four compute nodes, namely GL5, GL6, GL2, and MOX2. The HPC cluster is arranged as a tree, which means that the head node creates and distributes jobs on the compute nodes. A user can set the jobs to be executed, allocate resources to the jobs, retrieve statistics about job execution, and retrieve the output resulting from job execution. These commands are handled by the \texttt{slurmctld} daemon that runs on the head node. This daemon communicates with the \texttt{slurmd} daemons that run on each compute node. Each \texttt{slurmd} daemon is responsible to track job execution and report information to the head node. These statistics include the status of the node, information about job status, as well as job performance, and energy consumption. The head node records the information about job execution in a MySQL database. The interactions between the database and the \texttt{slurmctld} daemon are managed by the \texttt{slurmdbd} daemon that we omitted in Figure \ref{fig:hpc} for simplicity. The output resulting from jobs execution is stored into a shared folder that is accessible by all the nodes of the cluster. A user can collect the information stored in the database using the \texttt{sacct} command of SLURM and the data contained in the shared folder through SSH. 


The Green Lab includes about 20 smartphones and 3 tablets for executing experiments on \textit{mobile} applications, including several generations of Google Pixel phones, Samsung J7 phones, \etc. In addition, the Green Lab has 32 Raspberry Pi (of which, 2 Pi 5 Model B, 16 Pi 4 Model B, and 14 Pi 3 Model B) and 8 Arduino Nano to test embedded and IoT software applications. For experiments involving \textit{AI models}, in addition to using the cluster, the lab has also 3 Jetson Nano development kits. For \textit{Robotics} Systems, the Green Lab has 5 TurtleBot3 Burger robots (equipped with a camera module and current sensors) and a 4-degree of freedom robot arm (UARM Swift Pro, UFactory). Additional equipment includes \textit{wearables}, such as Samsung Gear S3, \textit{virtual reality} visors, \ie a Google Daydream, and a \textit{drone}, namely a DJI Tello.

\subsection{Measurement Infrastructure}

Green Lab users can track energy consumption and performance of the software running in the lab using both software energy profilers and hardware power meters. The cluster is equipped with two Rittal Power Distribution Unit (PDU) \cite{2023_rittal} and a Watts Up? Pro \cite{2023_wattsup} that profile the power requested by each node. A Rittal PDU is a power strip that provides reliable power distribution with energy monitoring and measurement for all devices connected to it. A Watts Up? Pro, instead, is a plug power meter that is placed between the device to profile and its power source. Both power meters provide an API to get the measurements. In addition, the Rittal PDU provides a web interface to configure the power strip and monitor the power required by the servers in real time. 
At the current state, the two Rittal PDU supply and monitor power of all the nodes of the cluster. 
We use the Watts Up? Pro to profile the power consumption of a node and also of other embedded systems, such as smartphones and Raspberry Pis. The Green Lab provides 3 Watts Up? Pro, including the one connected to the cluster. The Green Lab also features 3 Monsoon Power Monitor \cite{monsoon_2022}, which are high-frequency power meters. We mainly use them for experiments involving mobile applications but they can also be used to profile other embedded systems. A Monsoon can be plugged into a device through a USB port, a Main Channel consisting of a positive and a negative terminal, and a BNC connector referred as Auxiliary port in the Monsoon documentation. We use the USB port to profile USB-powered devices and the Main Channel for tracking the battery of our smartphones. The Main Channel can be connected to all devices exposing a positive and negative terminal, such as the pins of a Raspberry Pi. The power consumed by the Raspberry Pis and the Arduino Nano can be monitored also using an INA219 sensor. The INA219 \cite{ina_2023} is a current sensor that can be soldered on a board or connected to the board through a breadboard. The Green Lab comprises 5 INA219 at the moment.

The above-mentioned physical power meters record the energy consumption of the whole device while running a software application. A handy solution for fine-grained energy consumption measurements are software energy profilers \cite{cruz2021tools}. Software energy profilers avoid direct interaction with the hardware platform and allow more fine-grained measurements on the software. For example, at the process or code block level. Among the most used software profilers, there are \texttt{powerstat} \cite{2023_powerstat}, \texttt{perf} \cite{2023_perf}, and \texttt{Intel Power Gadget} \cite{2023_power_gadget}. All the mentioned tools exploit the Running Average Power Limit (RAPL) \cite{rapl2023} interface provided by Intel CPUs. At the moment, RAPL is used by most software energy profilers. RAPL estimates the energy consumed by a device according to a power model, which is proven to be accurate on certain architectures, and stores the estimate in a set of registers called model-specific registers (MSRs). SLURM, which we use for the HPC version of Green Lab, also uses RAPL to derive the energy consumed by the jobs executed in the cluster. However, not all nodes in our cluster support RAPL. The older ones, \ie GL3 and GL4, do not support RAPL. For this reason, GL4 is used only for dispatching jobs and record the measurements in the HPC configuration.

\subsection{Guidelines for Proper Use of Equipment}

\ins{It is crucial to provide proper instructions to students on how to use the tools and equipment provided for their projects; major misuses of the tools and equipment can affect the completion of their projects or can even lead to physical hardware damage. The definition of the experiment design during Assignment 2, see Section \ref{sec:project}, is followed by a phase in which the scientific assistants of the S2 Group interact directly with each group to provide them with support material and instruct them on how to use it. The groups are provided by the scientific assistants with credentials to access a limited set of resources, namely CPU cores, RAM, disk, and network bandwidth, on the servers according to the requirements of the experiment, as well as instructions on accessing them, usually through SSH.  Scientific assistants provide step-by-step guides for setting up embedded devices like Raspberry Pis, Virtual Reality Headsets, and educational robots, as well as instructions for maintenance. For example, we instruct students to always check and set the voltage provided by the power source (\eg the Monsoon power meter) to values that are not greater than those supported by the device to be powered (\eg 5V for the Raspberry Pi). The same applies to the physical power meters, \ie the Wattsup Power Monitor and the Rittal PDU. In addition, we provide students with material that could be helpful in better understanding the usage of the devices, such as links to the manual of the devices and online user discussions. The Green Lab course includes a lab, mentioned as LAB1 in Section \ref{sec:contents}, that covers the use of software tools to manage experiments, measure performance, and track energy consumption. During the lab, the instructor explains how to set up the tools and provides a live demonstration of how to use them for experimentation. The scientific assistants and the lecturers are available to support the students with technical problems and questions about the usage of measurement tools during the whole course. This is usually accomplished through open discussions on the web platform of the course.} 
\section{Open-Source Tools and Community of Learners}\label{sec:tools}

In the first three editions of the course, the development of a proper measurement pipeline for executing the experiment was an additional task for the students. However, we noticed that for projects that were focusing on the same technologies (\eg projects on the energy consumption of PostgreSQL databases or the front-end of mobile web apps) students tended to ``reinvent the wheel'' in their scripts in each of their projects. Examples of redundant scripts include tasks such as: defining the order of execution for experiment trials, shuffling experiment runs based on the considered subjects and treatments of the main factors, programmatically launching mobile apps, checking the status of smartphones, configurating energy profilers, \etc Even though those tasks were relatively exciting for students, they were very repetitive and error-prone; we saw an opportunity to save students' (and researchers') time here and we decided to move from ad-hoc development of measurement pipelines to the \textit{configuration} of measurement pipelines via \textit{dedicated tools}. 

So, in the months before the 2017/2018 edition of the course we decided to work on a first tool for the automatic execution of measurement-based experiments. To keep the scope of the tool under control, we decided to focus on experiments targeting mobile apps (both Web and native ones) running on Android devices (both smartphones and tablets). This tool is called \textbf{Android Runner}\weblink{https://github.com/S2-group/android-runner}~\cite{A_Mobile_2020}.
Android Runner is implemented as a set of Python modules, and as such it can run on any machine able to run Python code (\eg laptops, Raspberry Pis, servers). 
Given a JSON-based configuration of the  experiment, Android Runner is in charge of performing diagnostic checks, bringing up (energy) profilers, executing each run of the experiment according to a predetermined plan, collecting and aggregating measures from the profilers, \etc 
Android Runner communicates with the measured apps via Android Debug Bridge (ADB\weblink{https://developer.android.com/tools/adb}), the official Android command-line tool for interacting with Android devices. 
Android Runner is independent of the specific energy profilers (which are implemented as external plugins), allowing users to straightforwardly collect run-time measures via already-existing hardware/software profilers or by integrating their own third-party profiler; Android Runner allows users to use multiple profilers within a single experiment, \eg for collecting performance and energy measures at the same time (as we did in \cite{IST_2024}).
Moreover, Android Runner allows users to include their business logic as external Python scripts at specific points within the experiment execution (\eg before the beginning of the experiment, before or after each run, \etc); over the years, external scripts proved to be useful in many situations, \eg for setting up a local web proxy recording the traffic generated by the apps, for instrumenting the subjects of the experiment on the fly, for cleaning up the environment between runs, \etc

Over the years, and thanks to the lessons learned while working on Android Runner, we decided to develop other two tools for executing measurement-based experiments: (i) Robot Runner is dedicated to robotic systems and (ii) Experiment Runner, which is a generalization of the previous *-Runner tools, allowing students (and researchers in general) to configure measurement-based experiments in a technology-independent manner.

\textbf{Robot Runner}\weblink{https://github.com/S2-group/robot-runner}~\cite{ICSE_2021} follows the same principles as Android Runner, with the following main differences: (i) it communicates with experiment subjects via Robot Operating System (ROS\weblink{https://www.ros.org}) messages instead of Android ADB, (ii) the internal diagnostics checks are specific to ROS-based system, (iii) it is completely independent of the type and number of used subjects (either simulated or real), whereas Android Runner assumes to be interacting with a physical and always-available device, and (iv) the experiment configuration is defined in Python instead of JSON. About the latter point, the rationale for having a Python-based representation of the experiment (instead of JSON) is to allow users to flexibly define the configuration and the business logic of their experiments in a \textit{self-contained manner}~\cite{ICSE_2021}. This is aligned with the ROS ecosystem, where mission launch files for the robotic system can be defined in Python\weblink{https://docs.ros.org/en/iron/Tutorials/Intermediate/Launch/Creating-Launch-Files.html}. 

\textbf{Experiment Runner\weblink{https://github.com/S2-group/experiment-runner}} is our latest tool and it embodies several of the lessons learned from both Android and Robot Runner. Specifically, it allows users to (i) programmatically define experiments in terms of a so-called \textit{run table} containing experimental factors and their related treatments, subjects and objects, and their combination into trials, (ii) start, stop, pause, and resume experiments, (iii) run the experiment in automatic mode (without any interruptions) or semi-automatic mode (where the experiment waits for a user's action before at the end of a run before proceeding with the next one -- this is useful in cases where a manual action is needed in between runs), (iv) intuitively keep track of the progress of the experiment, (v) define custom Python callbacks that are invoked when specific events are raised during the execution of the experiment (similarly to Android and Robot Runner). During the execution of the experiment, the run table defined by the user is populated dynamically with the measures collected at each run, without requiring the user to manually aggregate them at the end of the experiment. Similarly to Android and Robot Runner, measures are collected via external profilers.        


\begin{figure}
   \centering
   \includegraphics[width=.9\linewidth]{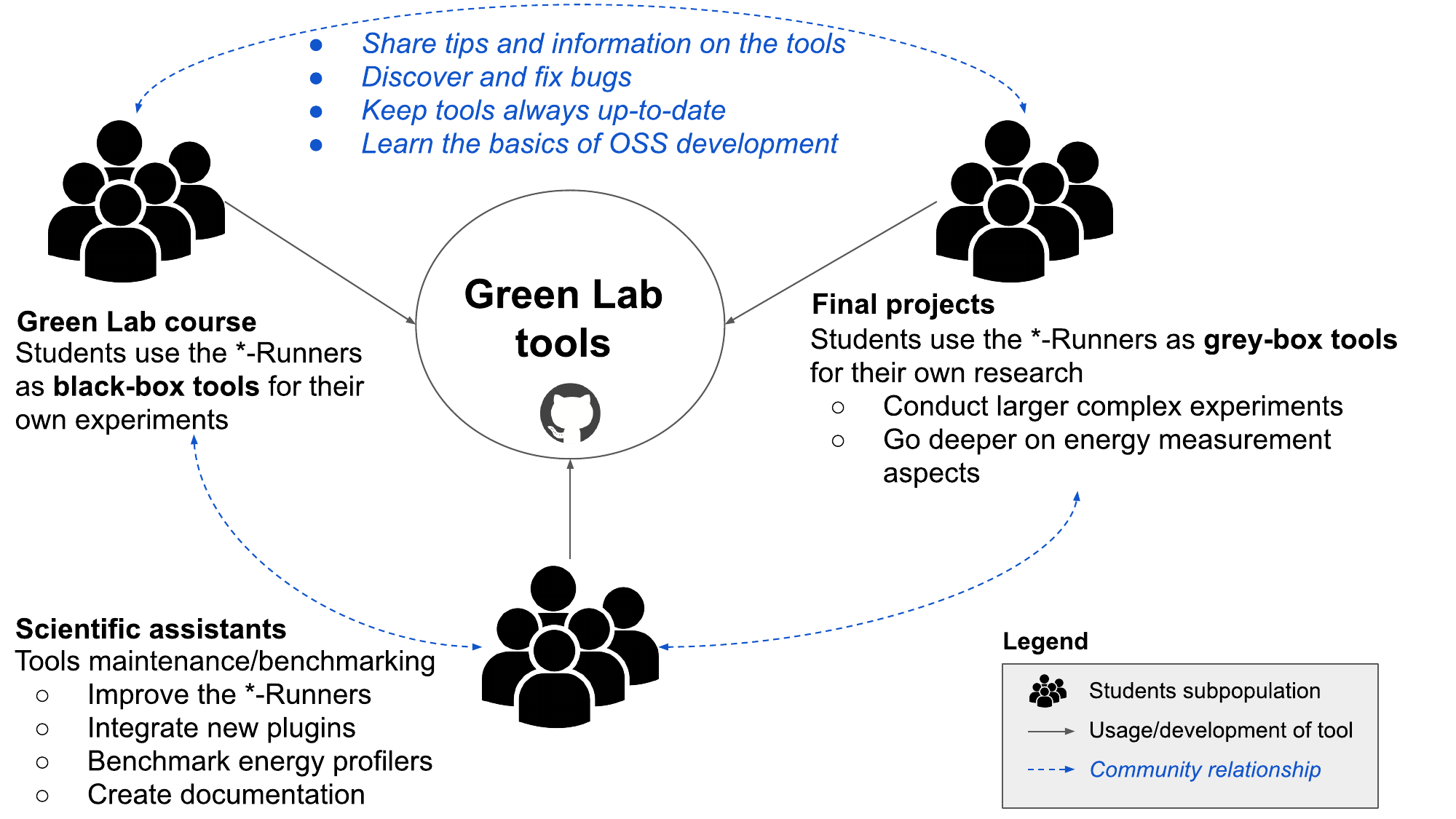}
   \caption{Green Lab community of learners}
   \label{fig:community}
\end{figure}

At the time of writing, the above-mentioned tools have been used to execute a total of 68 experiments within the Green Lab course and many others outside the scope of the course (\eg in experiments not linked to education or in students' final projects). As shown in Figure~\ref{fig:community}, the three above-mentioned tools are the center of a large \textbf{community of learners} composed of students attending the Green Lab course, students carrying out their final projects (both Bachelor and Master), researchers of the S2 research group, and students acting as scientific assistants within the S2 group.
Specifically, Green Lab students use the ``Runners'' as black-box tools for executing their own experiments for completing the project of the course; they are not required to know the internal details of the orchestration logic of the tools, but in almost all cases they integrate the tools with their own custom logic via external Python scripts.
Students carrying out the final projects use the Runners for executing their own experiments, which tend to be more complex and large than those of the Green Lab course. Also, they might go deeper on selected aspects of energy measurement, \eg by benchmarking new tools and profilers, benchmarking already-existing energy meters/profilers against each other, \etc   
Scientific assistants are generally Master students enrolled within the S2 research group all year long to support the development of the Runners; their responsibilities include the improvement of the tools, bug fixing, integration of new plugins and their benchmarking, and documenting the tools via videos, tutorials, examples, guides, and readmes.
Together with S2 researchers, these three sub-populations of students form a community of learners where individuals share tips and (less documented) information about the tools, discover and fix bugs in the tools, and keep the tools up-to-date. The mechanisms we use for synchronizing all those activities are exactly the same as those of \textit{collaborative software development}\weblink{https://www.linuxfoundation.org/blog/getting-started-with-open-source-development}. The source code of all Runners is publicly-available in their GitHub repositories, learners fork those repositories, develop and test their fixes or new plugins in their own forks, submit their own contributions to the upstream GitHub repository via pull requests, and receive feedback from S2 researchers and other learners and (if needed) make changes accordingly. The community is growing more and more -- as of today it counts more than 300 participants in total (past and present) -- and we believe it is the ideal setup for training students and young researchers to participate in collaborative software development, a skill that is deemed as a must-have in the job market for software engineers.         
\section{Success Stories}\label{sec:success_stories}

This section presents examples of recent relevant scientific studies that have been designed, conducted, and reported in the context of the Green Lab. All studies have been positively received by the software engineering research community and tended to receive primarily positive reviews by our peers. 

\ins{Before describing the success stories, it is important to elaborate on how we organize the transition from an education-oriented project developed by 5 students to a scientific publication able to compete in the Software Engineering research landscape. This transition is carried out in 4 phases: (i) projects supervision,  (ii) projects selection, (iii) authorship discussion, and (iv) writing and submission.} 

\noindent \ins{\textbf{Projects supervision}}. This is a long-term phase starting at the beginning of the course and ending with the final submission of students' projects. As already discussed, each project is split into 3 separate assignments, and instructors provide detailed feedback about each of those assignments (the mindset is to give provide a review similar to the ones done in top Software Engineering conferences). Via those 3 assessment moments instructors are able to (i) steer the projects towards research excellence (\eg avoiding known methodological pitfalls or choosing reliable energy measurement tools) and (ii) ensure that the projects remain on track and relevant to the Software Engineering community.   

\noindent \ins{\textbf{Projects selection}. The instructors select a subset of students' projects having the highest potential in terms of scientific value. This selection is carried out by the whole team of instructors and it is based on their overall experience in conducting controlled experiments; in this context, instructors identify and discuss the top projects in terms of (i) robustness of the technical aspects of the experiment, (ii) solidity of the design of the experiment, (iii) replicability of the experiment, (iv) potential scientific relevance and/or practical impact of the experiment, and (v) quality of the reporting of the experiment. The number of projects passing this selection phase range from 2 to 9 projects per edition of the course. The selection phase allows instructors to have a very good starting point for the scientific publication, considering also that the team of students invested more than 800 hours in studying the material and carrying out the project (the expected effort of each student is 168 hours in total for the whole course -- see Section \ref{sec:contents}).} 

\noindent \ins{\textbf{Authorship discussion}. 
The list of authors of the scientific publication is composed of students (always first in the author list) and one or more members of the teaching team (always last in the author list).
Firstly all students who worked on the project are asked to participate to the scientific publication, clarifying that the additional effort from their side is generally very minimal. Then, those students who accepted to participate in the scientific publication are included in the author list, and the others who did not\footnote{This is a very rare  case and it is primarily due to students having already a full schedule for the remainder of their academic year.} are included in the acknowledgment section of the paper. 
Then, one or more members of the teaching team (\ie the lecturers and all teaching assistants) is associated to the scientific publication. As previously discussed, in the first lecture of the course the lecturer presents a list of possible topics to students. Such list of topics is created and discussed collaboratively by the whole teaching team; each possible topic is associated to the member(s) of the teaching team who initially proposed it; they act as the \textit{owner} of the scientific publication and will drive the next phase. 
}

\noindent \ins{\textbf{Writing and submission}.
Students are asked to address all feedback provided in the context of their final assignment. Then, the owner(s) of the scientific publication takes care of (i) checking the final revision of the study provided by the student, (ii) rewriting parts that are generally more customary for a scientific publication (\eg expanding the introduction with a summary of the main results and contributions, revising the related work section, \etc), (iii) expanding and elaborating more in details the main implication of the study (\eg in the Discussion section), and (iv) forking the initial replication package of the study produced by the students, refining it (when needed), and anonymizing it (in case of double-blind submissions). Students remain available during all the previously-listed activities in case clarifications are needed. Finally, the scientific publication is revised by all authors and submitted by the owner. 
}

\subsection{Scientific Studies on Web Technologies}


\subsubsection{Impact of service workers on the energy efficiency of Web apps~\cite{Mobilesoft_2017}}

This was the first scientific study emerging directly from a Green Lab project and it was awarded with the Distinguished paper award at MOBILESoft 2017 (co-located with ICSE). The lessons learned and the various interactions with the students during this study led to the creation of Android Runner~\cite{A_Mobile_2020}. 

\begin{wrapfigure}[16]{r}{0.5\textwidth}
  \begin{center}
    \vspace{-1cm}
    \includegraphics[width=0.5\textwidth]{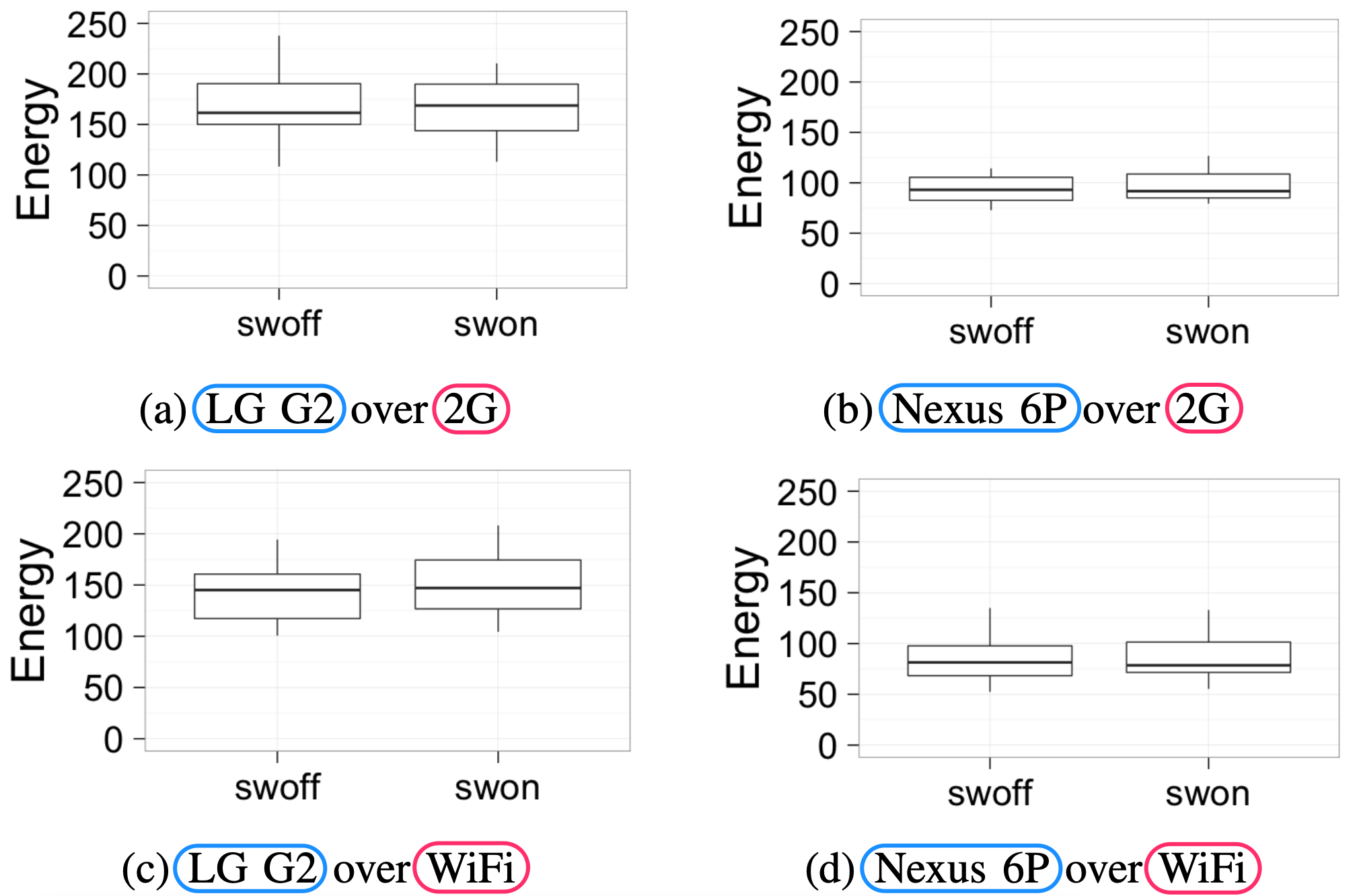}
    \vspace{-.9cm}
  \end{center}
  \caption{Energy consumption in Joules of the 7 Web apps with (swon) and without service workers (swoff) on different mobile devices (LG G2, Nexus 6P) and under different simulated network conditions (2G, WiFi) -- adapted from \cite{Mobilesoft_2017}}
  \label{fig:mobilesoft_2017}
\end{wrapfigure}
Service workers\footnote{\url{https://www.w3.org/TR/service-workers}} are a W3C standard providing APIs to allow Web developers to programmatically preload assets required by a web apps, cache data received from servers, subscribe to and receive push notifications, \etc 
The main motivation of this study is that service workers were advertised by several technology players as performance boosters, network savers, and providers of better user experience; however, despite they are additional code to be downloaded, parsed, and run by the browser, to that day nobody investigated the potential 
overhead in terms of energy consumption. 
As shown in Figure~\ref{fig:mobilesoft_2017}, this experiment had two main factors: the use of service workers and the type of (simulated) network
available (2G and WiFi). The subjects of this experiment are 7 third-party and publicly-available Web apps, \eg Washington Post, Ali Express, Wikipedia. We run the subjects on Google Chrome via two Android devices (LG G2 and Nexus 6P); the Android devices were used as blocking factor. 
The dependent variable of the study is the energy consumption of the Android devices, measured via the Trepn Power Profiler. The experiment did not provide statistically-significant evidence about the impact of service workers on energy consumption,
regardless of the network conditions; no interaction was
detected between the two main factors of the study.

This study is also relevant in the context of scientific integrity; indeed, it is a good example of the value of non-conclusive studies in terms of statistical significance, allowing students to understand the value of a scientific study independently of reaching statistical evidence.

\subsubsection{Correlation between performance scores and energy consumption of mobile Web apps~\cite{EASE_2020}}

This study has been carried out in collaboration with Greenspector, a French company specializing in environmentally sustainable digital transformation of organizations.
Greenspector contributed to the study by allowing our students to collect measures via their software-based energy profiler for Android devices. 

\begin{wrapfigure}[14]{r}{0.6\textwidth}
  \begin{center}
    \vspace{-1cm}
    \includegraphics[width=0.6\textwidth]{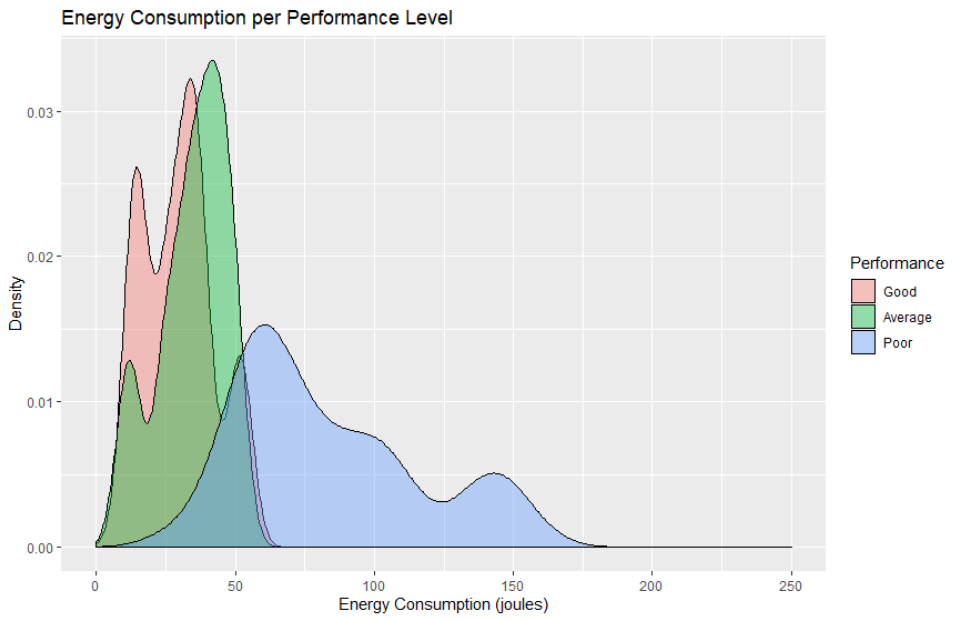}
    \vspace{-1cm}
  \end{center}
  \caption{Density of energy consumption among three performance levels  -- from \cite{EASE_2020}}
  \label{fig:ease_2020}
\end{wrapfigure}
The study is based on the observation that Web developers can use several tools for collecting performance metrics of mobile Web apps (\eg Google Lighthouse\weblink{https://developers.google.com/web/tools/lighthouse}), but similar ready-to-use tools are (still today) not available for energy metrics.
The goal of this study is to investigate whether the metrics produced by Google Lighthouse can be used as a proxy for energy consumption.
To answer this question, we conducted an experiment where 21 real
mobile web apps (\eg \texttt{apple.com}, \texttt{theguardian.com}, \etc) were analyzed in terms of their performance
level (via Google Lighthouse) and their energy consumption. 
After having collected the measurement data for 525 runs (21
web apps, 25 repetitions each), we statistically analyzed the correlation between the obtained performance metrics and energy
consumption, and carry out an effect size estimation.
The results of the study provided empirical evidence about a statistically significant negative correlation between performance levels and the energy consumption of mobile web apps, with medium to large effect sizes (see Figure~\ref{fig:ease_2020}).
In practical terms, this means that if a web app has a good
performance score on mobile devices, then developers can use such
a score as a low-cost alternative for preliminary insights about its energy consumption. 

This study attracted the attention of Web developers and it is used as one of the support resources of the current draft of the Web Sustainability Guidelines\weblink{https://w3c.github.io/sustyweb}, a specification providing recommendations for making websites and products more sustainable published by the W3C Sustainable Web Design Community Group.


\subsubsection{Performance and energy costs of ads and analytics in mobile Web apps~\cite{IST_2024}}
This study is published in the Information and Software Technology journal and it is the first journal publication emerging from a Green Lab project.

\begin{wrapfigure}[12]{r}{0.6\textwidth}
  \begin{center}
    \vspace{-1cm}
    \includegraphics[width=0.6\textwidth]{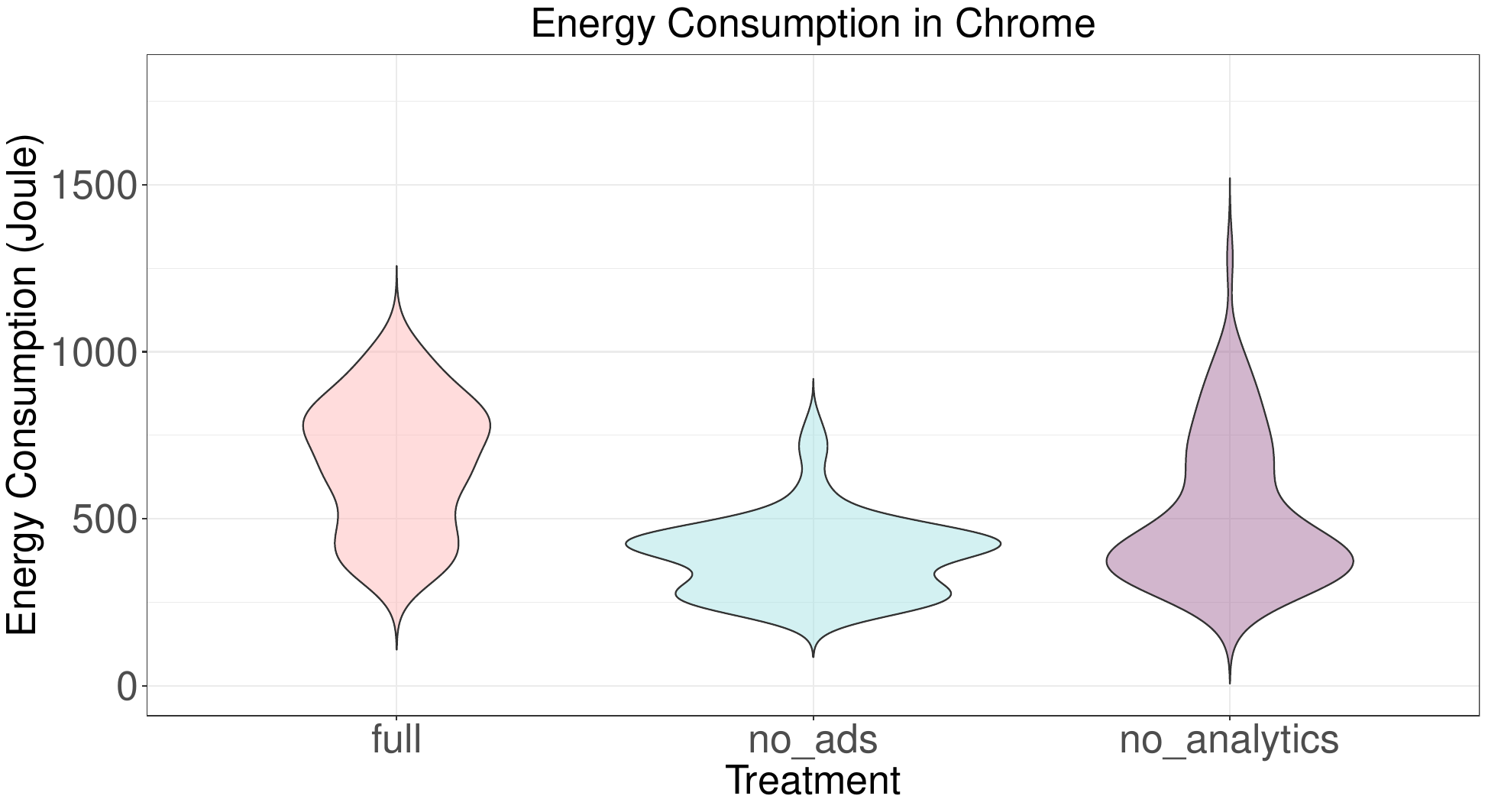}
    \vspace{-1cm}
  \end{center}
  \caption{Energy consumption of web apps (full, without ads, and without analytics)   -- from \cite{IST_2024}}
  \label{fig:ist_2024}
\end{wrapfigure}
This study targets ads and analytics in Web apps running on mobile devices.
The premise of the study is that ads and analytics have inherent costs due to them being additional software modules running in a Web app and making additional network requests.
Subsequently, more computing resources are used, potentially having an impact on the energy consumption and the performance of Web apps.
This study aims to analyse the performance and energy
overhead of ads and analytics on mobile web apps. The results of this
research are intended to provide developers, browser vendors, and
researchers with insights into the energy consumption and performance drain caused by running such software on consumer mobile devices.
Students sampled 9 popular web
apps containing both ads and analytics from the Tranco list~\cite{tranco}, a stable research-oriented list of the top one million most popular web applications.
For each web app, students obtained three versions: the full original web app, a version without ads, and a version without analytics. 
Then, each version of each web app is loaded multiple times on an Android device via
two different browsers, namely Google Chrome and Opera.
The energy consumed by the browser for loading the subjects is measured using the BatteryStats plugin of Android Runner. Two performance metrics are collected via the PerfumeJS plugin of Android Runner, namely (i) First Contentful Paint (FCP), and (ii) Full Page Load Time (FPLT). 
The main results of the study are:
(i) ads significantly impact the energy consumption of mobile web apps for both browsers, with a large effect size (see Figure~\ref{fig:ist_2024}); 
(ii) analytics have a significant impact on the energy consumption of Chrome, with a medium effect size (see Figure~\ref{fig:ist_2024}), but not on Opera; 
(iii) both ads and analytics do not significantly impact the FCP metric on both browsers; and (iv) both ads and analytics significantly impact the FPLT metric on both browsers, but with a small effect size.
This study provides evidence that both ads and analytics do have a significant impact on the
energy consumption and performance of mobile web apps. 
Web developers are advised to limit both ads and analytics in their mobile web apps to reduce their energy consumption and improve their loading time.

\subsection{Scientific Studies on Mobile Apps}


\subsubsection{Comparison of the energy consumption and performance of native Android Apps and their Web counterparts~\cite{MOBILESoft_2023}}

This study compares the quality of the native version of a mobile application against its web counterpart. This work has been (remotely) presented at MOBILESoft 2023 by a student, thus giving relevance to the commitment of the students and the quality of both the course and the lab. This study observes the problem mainly from a user perspective. If correctly informed, users may choose to use the native or the web version of an application to save battery life and for better performance. 
\begin{wrapfigure}[32]{r}{0.65\textwidth}
  \begin{center}
  \vspace{-1cm}
    \includegraphics[width=0.65\textwidth]{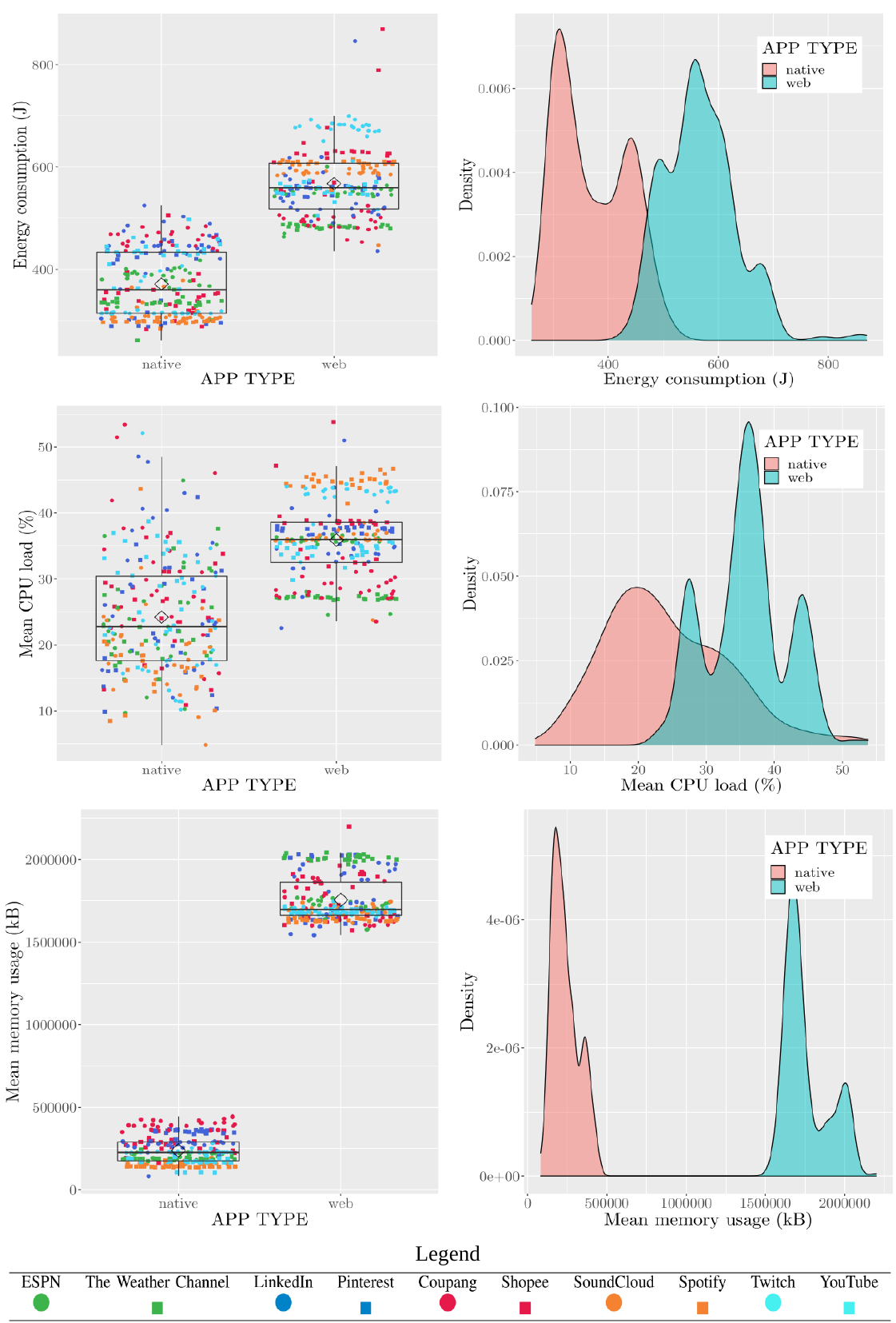}
    \vspace{-.5cm}
  \end{center}
  \caption{Energy consumption (top row), CPU load (middle row), and memory usage (bottom row) for 10 mobile applications -- adapted from \cite{MOBILESoft_2023}}
  \label{fig:mobilesoft_2023}
\end{wrapfigure}
%
Our daily lives have become more and more dependent on mobile applications. Music streaming platforms, digital newspapers, and social networks are just a few examples of widely-used mobile applications. On-demand content is usually provided through two versions of a mobile application: a native and a web version. Native applications are designed for a specific system and have direct access to the hardware resources of the system. For example, native applications can be specifically designed for Android and fully support the GPS, camera, and accelerometer of the device. The web version of the application, instead, is portable and implemented using web languages, \eg HTML, CSS, and JavaScript.  
A set of 10 well-known mobile applications is sampled from different categories, such as social networks and news. Each application is stressed with a typical usage scenario for that application. For example, the scenario executed to profile the news application of ESPN involves a user opening an article, scrolling down, and checking the next one. Both versions of the application are tested using default settings. A scenario is executed 25 times for each application, profiling the performance and the energy consumed per run. 
Figure \ref{fig:mobilesoft_2023} depicts the results of this study. The top row shows that the native apps tend to be more energy-efficient than their web counterparts. The most energy greedy applications are social media and e-commerce with Twitch as the most energy consuming application. The web version resulted in using considerably more CPU and memory, as shown by middle and bottom row of Figure \ref{fig:mobilesoft_2023}. Instead, there is no significant evidence that native applications have improved network traffic and frame time.

\subsubsection{Run-time efficiency of Android apps migrated to Kotlin~\cite{SCAM_2021}}\label{sec:scam_2021}

This study is the result of a Master thesis carried out using the Green Lab tools and its main characteristics is its \textit{mixed-method} research design involving a GitHub mining phase and a measurement-based experiment. 
The study was awarded with the Distinguished paper award at IEEE SCAM 2021 (co-located
with ICSME). 

\begin{figure}
  \begin{center}
    \vspace{-.5cm}
    \includegraphics[width=1\textwidth]{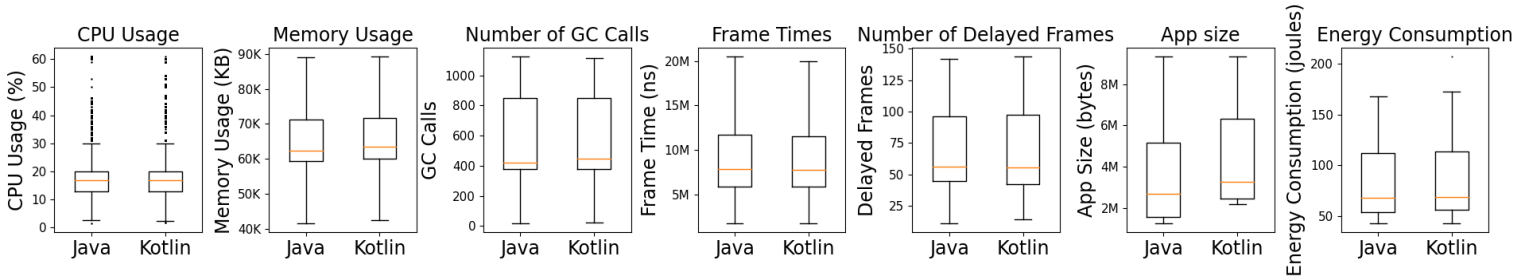}
    \vspace{-1cm}
  \end{center}
  \caption{Run-time efficiency of Java and Kotlin apps on Nexus 5 -- adapted from \cite{SCAM_2021}}
  \label{fig:scam_2021}
\end{figure}

\vspace{-.5cm}

This study targets Android apps developed using Kotlin. 
Specifically, we aimed at empirically assessing the impact of the \textit{migration} from Java to Kotlin on the run-time efficiency of Android apps. 
To achieve this goal, the student mined 7,972 GitHub repositories containing Android apps and identified 451 apps containing Kotlin code. By applying a cross-language clone detection technique~\cite{cheng2016mining}, they detected 62 commits that represent a full migration to Kotlin, while keeping the app
functionally equivalent. 
This mining phase confirmed that most open-source Android apps either fully migrated to Kotlin ($>$90\% Kotlin code) or contained low portions of Kotlin code ($<$10\%).
After the mining phase, a sample of 10 apps that fully migrated
to Kotlin have been selected and then the measurement-based experiment has been executed to
compare their Java and Kotlin versions. 
The study considered six run-time efficiency metrics: CPU usage, memory usage, number of calls to the garbage collector, frame times, app size, energy consumption. 
The experiment has been orchestrated via Android Runner and targeted two smartphones. 
By referring to Figure~\ref{fig:scam_2021}, the results of the experiment highlight that migrating to Kotlin has a statistically significant impact on CPU usage,
memory usage, and render duration of frames (albeit with a
negligible effect size), whereas it does not impact significantly
the number of calls to the garbage collector, the number of
delayed frames, app size, and energy consumption.
This study provides evidence that \textit{developers can migrate their
Android apps to Kotlin and expect comparable efficiency at
runtime}. 

\subsubsection{Evolution of Kotlin apps in terms of energy consumption~\cite{ICT4S_2023}}

This a follow-up of the study described in Section~\ref{sec:scam_2021}. In this case, instead of looking at apps migrated to Kotlin, a team of Green Lab students performed a longitudinal analysis of the energy consumption of 3 selected Kotlin apps.
The goal of this study is to empirically assess how the energy consumption of Kotlin applications evolves over time. Students first rigorously selected three open-source Kotlin applications via well-defined inclusion and exclusion
criteria and a quality assessment procedure. 
Then, they cloned the GitHub repository of each app and selected a total of 57 combined releases of those apps. 
\begin{wrapfigure}[16]{r}{0.6\textwidth}
  \begin{center}
  \vspace{-1cm}
    \includegraphics[width=0.6\textwidth]{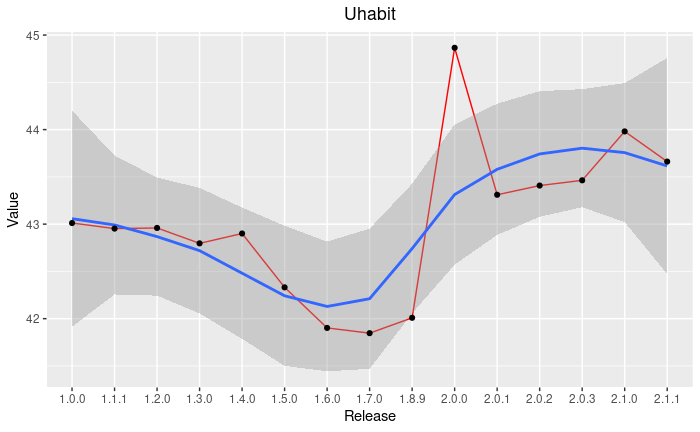}
    \vspace{-1cm}
  \end{center}
  \caption{Energy consumed by the Uhabits app across its releases in J (red line = raw energy consumption; blue line = LOESS smoothed consumption; grey area =
confidence interval 0.95) -- from \cite{ICT4S_2023}}
  \label{fig:ict4s_2023}
\end{wrapfigure}
The energy consumption of each of those 57 releases is measured using Android Runner and the collected measures are statistically analysed in order to
identify energy spikes and drops (see Figure~\ref{fig:ict4s_2023} for an example). Finally, source code changes occurring in ``energy-relevant'' releases are manually analyzed to identify possible causes for the observed energy fluctuations.
The study confirms that the energy consumption of Kotlin apps generally follows a growing
trend along releases, and in all apps significant energy
spikes and drops have been detected. Among the major factors that impacted those
fluctuations, we noted that those co-occurred with OS upgrades, the release of new
app features, the usage of poorly chosen design patterns and libraries, the injection of UI-related issues, and unstable app versions.

\subsection{Scientific Studies in other Domains}
\subsubsection{On the Energy Consumption and Performance of WebAssembly Binaries across Programming Languages and Runtimes in IoT~\cite{EASE_2023}}

This work represents the first project regarding IoT in the Green Lab and extends the work presented at EASE 2022 \cite{EASE_2022_web_assembly}. This study analyses the potential of Web Assembly (WASM) \cite{sletten2021webassembly} in another application domain, i.e. IoT. 
One of the students, who also authored the paper, presented the work at EASE 2023 in Oulu, giving the right coronation of effort, as well as recognition and visibility to the students involved in the project and the Green Lab.

WASM allows software written using non-web programming languages, such as C++ and Rust, to be executed inside a web browser. Modern browsers can load WASM modules through JavaScript and execute them through a separate browser sandboxed environment. WASM showed potential for increased security, portability, and close-to-native performance. For this reason, WASM represents a good solution for fast deployable and updatable software for IoT devices. IoT involves running software on a plethora of resource-constrained devices. The devices of an IoT system can vary from a few to thousands and need to be regularly updated to add new features and fix bugs. The adoption of WASM on IoT devices is achieved by using runtime environments. A runtime environment works as an interface between a WASM executable and the underlying system and provides basic system-related services, such as I/O operations.

In this study, we investigated the impact of different programming languages and runtime environments on the performance and energy consumption of WASM binaries in the context of IoT systems. We selected four different programming languages, namely Rust, Go, JavaScript, and C, and two different WASM runtime environments, \ie Wasmer, and Wasmtime. The programming languages were evaluated using three software applications taken from the Computer Language Benchmark Game (CLBG). We considered the implementation of each software in each programming language. Each implementation was executed using Wasmer and Wasmtime on a Raspberry Pi 3 Model B. The execution of the experiment resulted in 24 trials (4 programming languages x 3 benchmarks x 2 runtime environments), each one executed 10 times.

\begin{figure}
    \centering
    \includegraphics[width=\linewidth]{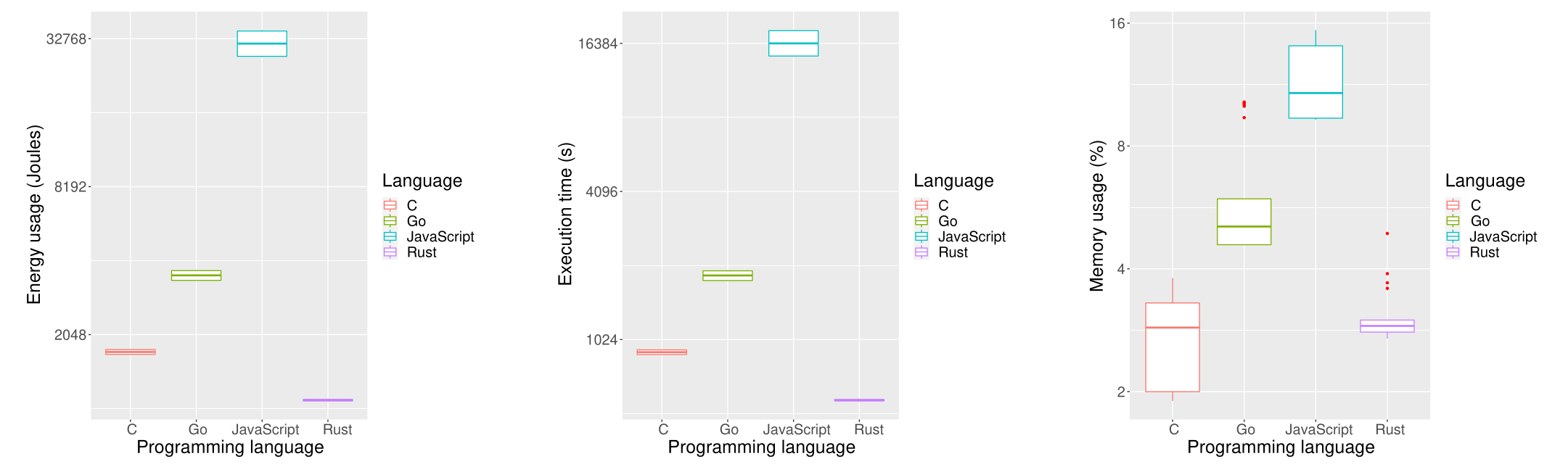}
    \caption{Energy consumption and performance of WebAssembly binaries across programming languages -- adapted from \cite{EASE_2023}}
    \label{fig:ease2023}
\end{figure}

Rust is the most energy- and time-efficient language, while C is the most memory-efficient (see Figure~\ref{fig:ease2023}). Instead, JavaScript resulted to be the most resource-demanding language. From the analysis of runtime environments, Wasmer is found to be more efficient than Wasmtime in terms of performance and power consumption. We study also the interaction between the runtime envinroment and programming language that showed energy and time efficiency for all programming languages running using Wasmer. For memory usage, instead, the executions through Wasmer resulted worse than those on Wasmtime.

\subsubsection{Computation offloading for ground robotic systems communicating over WiFi – an empirical exploration on performance and energy trade-offs~\cite{EMSE_2023}}
This study shows how we exploited the Green Lab to perform experiments in the area of robotics software. We set up a test environment for a TurtleBot3 Burger that moves autonomously in space. This environment is used to evaluate the impact of different offloading strategies on energy consumption and the performance of the robot by diversifying the tasks performed by the robot.

\begin{table}
\centering
\caption{Task parameter configurations -- adapted from \cite{EMSE_2023}.}
\label{tab:configurations}
\begin{tabular}{|p{3cm}|p{3cm}|p{1.6cm}|p{1.6cm}|p{1.6cm}|}
\hline \textbf{Factor}             & \textbf{Parameter}           & \textbf{Config. 1}   & \textbf{Config. 2}   & \textbf{Config. 3 }   \\ \hline
Object Recognition & image resolution    & 640x480px & 640x480px & 1280x960px \\ 
                   & image frame rate    & 20fps     & 20fps     & 60fps      \\ \hline
SLAM               & number of particles & -         & 5         & 30         \\ 
                   & temporal updates    & -         & off       & on         \\ \hline
Navigation         & velocity samples    & 10x20     & 10x20     & 20x40      \\ 
                   & simulation time     & 1.5s      & 1.5s      & 3s         \\ 
                   & map                 & know      & unknown   & unknown  \\ \hline 
\end{tabular}
\end{table}


Robotics system is known to be resource-constrained, which means resource utilization greatly impacts its operational time. For example, battery-powered robots need to optimize their energy consumption to ensure that they have enough power to accomplish their mission. To overcome hardware limitations, computation-intensive tasks can be offloaded to the Cloud. However, the integration of offloading mechanics within robotics software increases network utilization, which may reduce the quality of service and raise the energy consumption of the network.

This study empirically assesses the impact of offloading strategies on the performance and energy consumption of autonomous driving robots. The assessment involved two experiments. In the first experiment, our goal was to characterize offloading strategies and their impact on energy consumption and performance of the robot, while in the second, we analyzed the impact of the offloaded task parameter values on the same metrics. The robot subject to the experiment is a TurtleBot3 Burger~\cite{amsters2020turtlebot}, which performs three tasks: mapping (SLAM), navigation, and object recognition. These tasks are implemented using ROS. During the mapping task, the robot builds a map of the (unknown) environment in which it operates and localizes itself in the map. 
The navigation task, on the other hand, enables the autonomous driving of the robot to a goal location on a map, while object recognition consists of the identification of objects on the map by exploiting robot cameras. The tasks have different parameters, such as image resolution and frame rate for object recognition and the map for the navigation task that can be known or unknown. We observed the robot operating using different values of these parameters, referred in the paper as a configuration. Task parameter configurations considered in this study are described in Table \ref{tab:configurations}. The configurations have different magnitude, for example Conf. 3 considers the highest value for each parameter. The TurtleBot3 is observed operating in a test environment. For each task, we performed a mission with and without offloading the selected task to a remote PC. In the first experiment, we launched the robot with Conf. 1 and Conf. 2, which differ in the parameter on map knowledge, while in the second, we executed the robot with Conf. 3. Both experiments involved nearly 100 trials with a total experimentation time around 23 hours.

The results show that offloading reduces processing time for computationally intensive tasks, \ie object recognition. There were no significant differences in performance on other tasks. The experiment highlights increasing latency due to higher network utilization. The optimization observed for processing time reflects energy consumption, which is reduced when offloading compute-intensive tasks. The analysis of task parameters shows their configuration has a critical impact on performance and energy consumption. Thus, tuning the parameters of ROS tasks is critical to achieving energy and performance efficiency in robotics systems.

\subsubsection{Mining the ROS ecosystem for Green Architectural Tactics in Robotics and an Empirical Evaluation~\cite{MSR_2021_architectural_tactics}}

In this study, we exploited the Green Lab to validate the results obtained after obtaining four architectural tactics for energy-efficient robotics software from the literature. The Green Lab, in this case, proved effective in transferring and validating knowledge about software engineering practices to real-world contexts.  

Software is becoming central in robotics systems. For example, NASA uses more control software on its new exploration missions than it has on all of its previous missions combined. However, at present, robotic software is still poorly engineered, despite its increasing complexity and size. In addition, robotics software design focuses only on performance and functional aspects, leaving important issues such as energy consumption out of the design process. The integration of energy consumption analysis at design time would bring enormous benefits to the quality of the final system. Battery-powered robots could adopt strategies at run time to consume less energy and thus increase operational time.

This study identifies and empirically evaluates architectural tactics for robotics systems. The architectural tactics are identified by mining ROS-related data from open-source software repositories and Stack Overflow (SO). The collected dataset includes discussions on SO, GitHub pull requests, commit messages, and source code. After collecting the dataset, 
it is processed using a set of keywords related to energy concerns (\eg green, battery, power). Resulting data are manually validated by the researchers, which remove false positive and keep data concerning software architecture. The last filter applied to the collected data is a thematic analysis, in which the researcher manually grouped energy-related software architecture design decisions, \eg usage of low power mode, and threshold-based mechanisms. This process resulted in four architectural tactics for energy-efficient robotics software:

\begin{itemize}
    \item \textit{EE1 - Limit Task:} robot tasks, such as video streaming or robot navigation, can be limited when their energy consumption reaches a critical level.
    \item \textit{EE2 - Disable Hardware:} robot operation time can be extended by reducing unnecessary usage of hardware resources.
    \item \textit{EE3 - Energy-Aware Sampling:} sensors sampling frequency impacts the energy consumption of the robot. The sampling frequency of the sensors can be adjusted based on the energy level of the robot.
    \item \textit{EE4 - On-Demand Components:} robots components should be started only when needed and shut down when they are not required. This tactic saves energy consumed by components when not in use.
\end{itemize}

The tactics are empirically validated on a TurtleBot3 Burger \cite{amsters2020turtlebot}. The study evaluates the energy consumed by robotics when applying the tactics across different robot movement strategies and environments. The evaluation of the tactics involves a baseline (B) in which no tactics are applied and a combined treatment that involves the application of all the tactics simultaneously (C). The environments considered have obstacles or not, \ie empty or cluttered, while the movement strategies consist of no motion, random movement in the environment, \ie autonomous strategy, and following a prefixed plan, namely sweep.  

\begin{wrapfigure}[15]{r}{0.55\textwidth}
    \centering
    \vspace{-.7cm}
    \includegraphics[width=\linewidth]{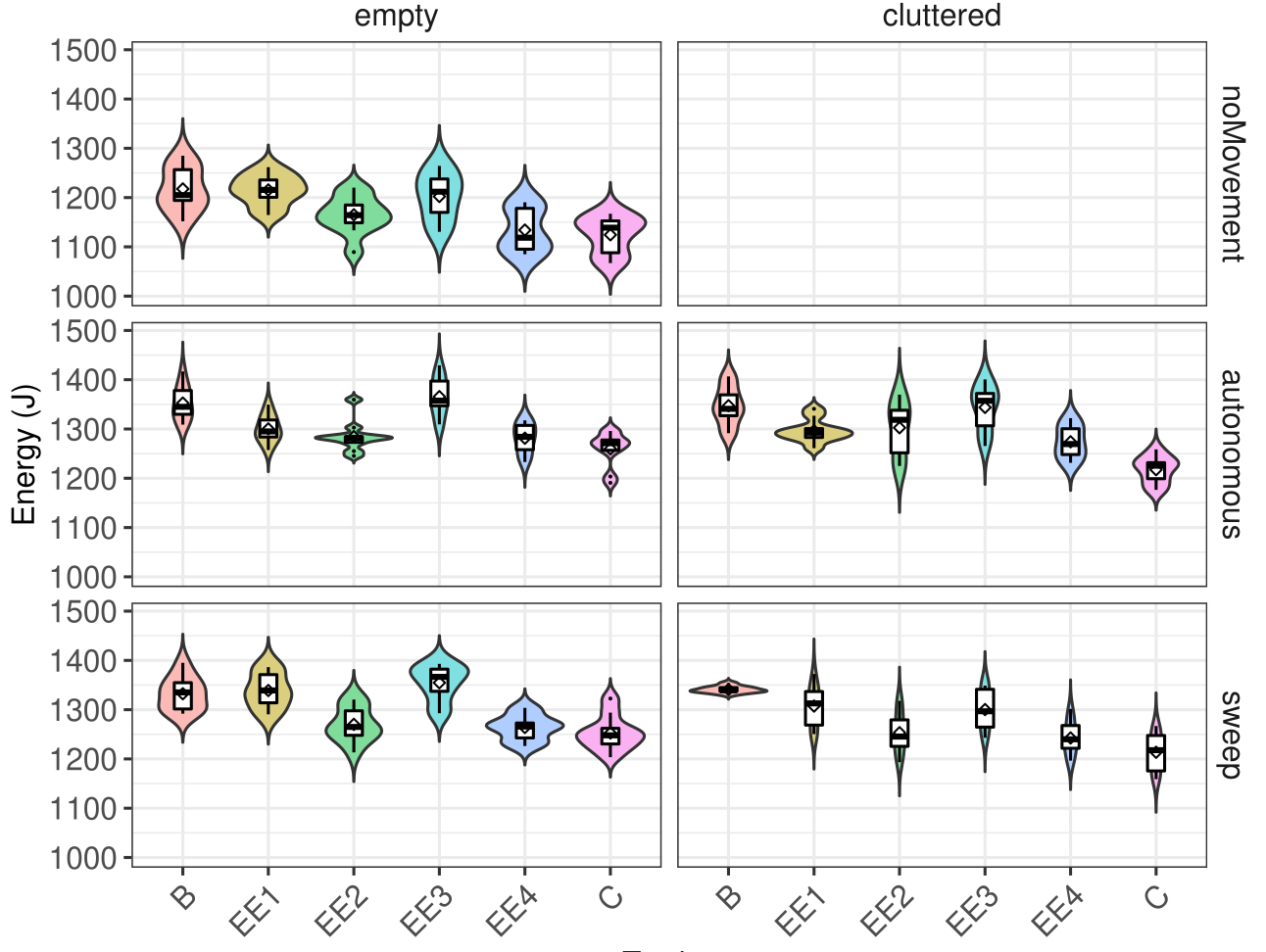}
    \vspace{-.5cm}
    \caption{Energy consumption across all movements strategies and environments -- from \cite{MSR_2021_architectural_tactics}.}
    \label{fig:msr2021}
\end{wrapfigure}

The results show that the application of the tactics improves energy efficiency. Moreover, combining all tactics results in greater energy efficiency than any single strategy in isolation. As shown in Figure \ref{fig:msr2021}, the energy consumed by the robot is highly influenced by the movements of the robot and the physical environment in which the robot operates. This study highlights a fundamental problem for both researchers and robotics software developers: energy consumption. The results of the mining phase confirm that energy is infrequently discussed by roboticists. Moreover, this work represents a starting point for researchers who want to define and refine further architectural tactics for robotics software, which are proven to be beneficial for robots' energy efficiency.

\section{Lessons Learned and Recommendations}\label{sec:lessons_learned}

The design of the Green Lab course described in Section~\ref{sec:course_design} is the result of ten years of experience in the delivery of the course and can be reused as a basis for delivering similar courses in other universities. We reached this stage iteratively. Over the years, we piloted several alternative solutions and various educational components and refined the content of the course edition by edition. This resulted in an accumulation of a number of lessons learned that we would like to share with the community. We report such lessons learned and make recommendations for instructors in the remainder of this section.

\subsection{General recommendations}

First of all, we learned that it is fundamental to \textbf{clarify expectations to students from day zero}. The strong research orientation of the course might not be common in some universities, and students might have a more didactic attitude concerning the course, instead of aiming to achieve scientific goals and solving open problems. In other words, students might be caught off guard when their project is evaluated with a critical eye from the instructors. As a solution, in the first lecture of the course, the instructors not only introduce the course (in terms of covered contents, learning objectives, assessment methods, and practical aspects), but also explicitly inform students that the teams whose project is well-reported and scientifically valid will be invited to participate in a scientific publication based on their project. In addition, as homework for the first lecture, students are asked to watch a video where one of the instructors presents the results of an experiment emerging from the course (specifically, it is about the results of \cite{Mobilesoft_2017}).     

Related to the previous point, \textbf{instructors should set high stakes for those teams that want to ``chase the stars", but they shall still be open to teams that just want to pass the course}. As the instructors say in the first lecture, ``this course is about opportunities''; however, it is important that students are not \textit{required} to catch such opportunities. In each edition of the course, there have always been teams that just wanted to pass the course with a good grade and learn just enough about ESE methods, without spending extra effort for the scientific publication. We believe that this is fair to students and those students shall still be able to get the maximum grade for their project. At the end of the day not all of our students are aiming to do a PhD in Empirical Software Engineering and, as instructors, we must accept it. 

Instructors should \textbf{keep the scientific publication not central to the project and value more students' learning than the scientific output}. This point is important to do not create too much pressure for students and it should be considered as a good extra-reward for the students.  
Concretely, during the first lecture students are asked to form their teams as soon as possible (recall that the deadline for the first assignment is generally at the end of the second week of the course) and each team is also asked to indicate if they are interested in contributing to a scientific publication based on their project (in the form we also allow for a \textit{``Maybe (if time allows)''} intermediate level).
In any case, instructors' feedback and grading rubrics are exactly the same for all teams. 
Also, in previous editions of the course, it happened also that a team was not willing to pursue the scientific publication based on their project; in those (rare) cases, the instructors did not proceed with the submission. 


In some past editions of the course, students raised concerns about the usefulness of the labs. We learned that in those cases the labs were (i) focussing primarily on the basics of the tools and R and (ii) less interactive than in the other years (\ie the teaching assistant was just presenting with slides and briefly carried out some practical tasks). 
The lesson learned here is that \textbf{labs shall be carefully planned in such a way that students can work on practical aspects related to their projects}. Higher lab interactivity can be achieved by: (i) asking students to compute basic summary statistics on real data coming from previous projects, (ii) creating scripts about the contents of the labs beforehand and making sure that they contain tasks for students, and (iii) investing time in educating the teaching assistants about the technical aspects of the lab (more technically-confident teaching assistants generally gave better labs).  

Designing and running a course like the Green Lab takes a considerable amount of time. This investment must pay off for instructors. First of all, a course like the Green Lab is a \textbf{good opportunity for researchers to carry out research in the context of their education activities}; in a way, running the course is not a purely teaching activity, but it is a way for researchers to have education and research \textit{intersect}, allowing instructors to continue doing their research activities in the weeks of the course (instead of putting them on hold). 
Moreover, by design, students who attend the Green Lab have the fundamentals of ESE methods in general (independently of the energy aspects). This means that those students are already able to design and conduct an empirical study in the context of, \eg their final project/thesis. Another strategy for capitalizing on this investment is to advertise the availability of further projects and theses during the course, allowing the instructors to have a \textbf{continuous research-oriented relationship with students even after the course}. For example, at VU Amsterdam there are two additional research-oriented courses that students can put in their study plan (for a total of 18 credits) and a 30-credit final Master project; if a researcher and a student agree to have the Green Lab, those two courses, and the final project on the same research topic, they will work together for a total of 54 credits, which is the equivalent of more than 8 months full-time. As such, this creates an opportunity for both the student and the researcher to design a more meaningful and well-structured research project than those carried out in more compact time frames.         

\ins{Over the years we also noted that courses like the Green Lab can be used as a good training environment for discussing \textbf{ethical aspects} related to empirical software engineering and IT in general. First of all, by directly targeting the energy efficiency of software, issues related to the carbon footprint of IT are inherently central in all lectures of the course; this is also by design since the course is given as a module of the Software Engineering and Green IT Master program and students enrolled in such program (i) expect this kind of discussions and (ii) have already a certain sensibility towards themes related to green software and sustainability. Moreover, by design the course touches upon aspects related to \textit{scientific integrity} and allows students to get a hands-on experience about the importance of (i) the replicability/reproducibility of scientific experiments (\eg some projects are about replicating an already-published study starting from its replication package and all student teams need to produce a complete replication package for their own projects) and (ii) properly reporting scientific studies (\eg all teams need to honestly describe the threats to the validity of their experiment and to follow empirical best practice when writing the report of their project).
During the three project assignments, the teaching team checks the ethical aspects of the experiments to ensure the reliability of the results and the claims derived from them. As the students evaluate the energy consumption of real software, their experiment may involve subjects and produce results that are sensitive to the stakeholders of the research, such as IT software developers, users, and IT companies.}

As reported in Section~\ref{sec:educational_context}, the duration of the Green Lab course is two months. We are aware that courses in other universities can have a longer duration. Below we indicate \textbf{how a course like the Green Lab might be expanded to longer courses}:
\begin{itemize}
    \item in-depth coverage of other ESE research methods (\eg repository mining, qualitative survey, systematic literature review) in addition to controlled experiments and integration of such research methods into students' projects;
    \item expanding the (currently single) lecture on techniques for developing energy-efficient software, possibly integrating it with intermediate experiments providing evidence about the discussed techniques;
    \item simulate a scientific conference by (i) setting up a reviewing system where students peer-review and discuss other teams' projects and (ii) teams present their results at the end of the course;
    \item organize ``replication sessions'' where teams carry out an additional experiment focussing on replicating (and learning from) an experiment carried out by other research groups in the world (this avoids the bubble effect where students are primarily exposed to experiments of the S2 group at VU Amsterdam);
    \item organize an intermediate hackathon on a selected common case (this might give a bonus to the winners of the hackathon);
    \item carry out more hands-on labs and in-depth labs on various measurement tools (and different domains -- IoT, mobile apps, servers, \etc).
\end{itemize}

\subsection{Recommendations about students' projects}

On top of experience and knowledge about ESE research methods, \textbf{the teaching team must have at least a basic knowledge of the technical/technological aspects of students' projects}. Given the relatively tight schedule of the course, the teaching team must be able to provide continuous (and timely) support to teams having technical issues during the course.
Possible strategies to achieve this include: (i) involve the scientific assistants mentioned in Section~\ref{sec:tools} as teaching assistants within the course (in this way it is the same people maintaining the tools supporting students in their usage), (ii) train teaching assistants on the technical aspects of the projects before the beginning of the course, \eg by executing small-scale replications of previous experiments, or (iii) enroll teaching assistants based on their technical skills and assign to them only projects related technologies related to such skills. 

It is suggested to \textbf{create multiple project tracks for each edition of the course}, roughly about double the number of student teams. In the first edition of the Green Lab, we had only two teams of students; so it was natural to have only two project tracks for the whole course. Differently, in the last edition of the course we had 19 teams of students, and having only two project tracks could have raised the likeliness of fraud among students and it could have made the assessment process extremely boring and repetitive for the teaching team. Thus, we opted to provide a large number of project tracks to students (\ie 30 individual tracks). Those project tracks touched upon different technologies (\eg Android APIs, data compression algorithms, large language models), platforms (\eg Web apps, VR apps, IoT, HPC), and types of research methods (\eg longitudinal studies, correlation studies, observational studies). When students register their team, they also provide their level of preference for each proposed project track, and then it is the teaching team that assigns the track to each team accordingly. This proved successful since students could choose for their project those tracks that best fit their skills and wishes. It is important to note that, in order to avoid assigning low-voted tracks, \textit{it is possible to assign the same project track to different teams}. If on one side this is a risk for the publication potential for the track (\ie only one team in principle will be able to publish its project), on the other side it creates safe competition among teams; also, it is up to the teaching team to propose slight variations to the redundant tracks in order to make them complementary (and thus equally valuable). As an example, our study on WebAssembly binaries across programming languages and runtimes~\cite{EASE_2022_web_assembly} has been performed also on Android.   

Creating multiple project tracks every year is a time-consuming and difficult endeavor. So, it is suggested to potential instructors to \textbf{keep a log of potentially-valid projects all year long}. Based on our experience, having a textual file with a 2-liner for each potential project track and some support links is more than enough. Then, at the beginning of the course, the teaching team meets, distributes the tracks among each other, and expands each 2-liner into one or two slides for each project track; those slides are then used in the first lecture to pitch project tracks to students. The list of potentially-valid projects proved also to be a useful overview of interesting research directions in the area of energy-efficient software, independently of the Green Lab course. Examples of sources we recurrently use for potentially-valid projects include (i) social media accounts of key tech engineers (\eg Addy Osmani for Web performance), (ii) engineering blogs of relevant tech companies (\eg Dropbox Tech Blog\weblink{https://dropbox.tech}),  (iii) websites focussing on software development and technological advances like InfoQ\weblink{https://www.infoq.com} and Ars Technica\weblink{https://arstechnica.com}, (iv) scientific literature in software engineering conferences, (v) developers' discussion forums like Stack Overflow\weblink{https://stackoverflow.com} in general or ROS Discourse\weblink{https://discourse.ros.org/} for robotics, and of course (v) daily discussions with peers and colleagues.

\subsection{Recommendations on tools and equipment}

\ins{Before delving into the details about tools and equipment,  we reiterate that the Green Lab course is designed as a \textbf{technology-independent course}. In other words, the learning objectives of the course, its structure, assessment procedure, and the majority of lecture contents are exclusively about energy efficiency and ESE research methods, and not bound to any specific technology (\eg Android, microservices, IoT) or application domain (\eg finance, social media). 
The only technology-specific components of the course are the following: (i) the topics of students' projects, (ii) the first practical session on the lab environment, tools, and devices, and (iii) the examples the instructor uses during the lectures. Being technology-independent is important for us (and future instructors willing to reproduce the course) since in this way the course does not need to be fully redesigned over the years. Being technology-independent, the course also allows students to acquire and apply ESE principles in other contexts, such as their final project (which might not even be about energy efficiency) or in their professional careers. Instructors willing to replicate the course in their own departments can reuse the course material as it is, with only minor adaptations, depending on available equipment  (which might even be students' laptops). In any case, \textbf{we suggest instructors to build an inventory of the hardware (and software) equipment available in their department before running an edition of the Green Lab course}. In this way, students can still measure energy consumption in a correct and reliable manner, without the need to buy expensive hardware. This is also a good opportunity for collaborating with other research groups within a department/university, for example by carrying out controlled experiments on the energy efficiency of software in the context of, \eg robotics, high-performance computing infrastructures, bioinformatics.}

In case of a large number of students (\eg above 50), \textbf{a good strategy to scale is to fix the technologies and used tools considered in the course}. For example, Android Runner was used by all students' groups for their Green Lab projects starting from 2017/2018 until 2022/2023. This allows (i) teaching assistants to save considerable effort in preparing for labs and supporting students in case of technical issues and (ii) the teaching team to assess students' projects more efficiently. However, it is important to note that this also comes at the cost of flexibility when defining students' projects, thus potentially impacting the research output of the course. For example, in the years when we were using exclusively Android Runner, all students' projects were about mobile (Web) apps on Android. Moreover, we often reuse technologies that are proven to be effective for specific domains, saving time and resources when preparing experiments. In the last edition of the course, we decided to go beyond Android and cover multiple technologies and platforms again; this was possible thanks to the possibility of financially supporting three teaching assistants dedicated to the course. Besides using our open-source tools, such as Experiment Runner and Android Runner, we usually integrate, for example, Prometheus~\cite{turnbull2018monitoring} and Grafana~\cite{chakraborty2021grafana} within experiments involving microservice-based applications. Prometheus and Grafana are open-source tools enabling monitoring and visualization of a set of metrics, such as CPU utilization and energy consumption. 

\textbf{Tools pay off, but they require continuous maintenance, otherwise, they fire back}. Over the years we learned that just having a tool like Android Runner is not enough to ensure a successful execution of experiments. For example, in previous years the Android Runner tool was tested on a relatively low number of smartphones and students using it on their own devices were having several compatibility problems; in those years we received some negative feedback from students about the fact that they were spending a lot of time in fixing issues with the tool, rather than learning how to conduct experiments on energy-efficient software. As a solution, we decided to allocate one teaching assistant for one day a week for the whole academic year whose main responsibilities are to (i) fix bugs in the tools used in the Green Lab, (ii) support students having issues with those tools, and (iii) implementing new features in the tools as new needs arise. As a result, after the injection of this new resource, we did not receive any negative feedback about tools in course evaluations, and students were able to focus on the main learning objectives of the course.    

Green Lab equipment is shared between several users, mostly students and researchers. The equipment is continuously maintained to avoid being accidentally damaged after usage. 
\textbf{Students shall be carefully instructed about equipment usage, both hardware and software}. For example, we usually use a Monsoon power meter, described in Section \ref{sec:equipment}, for experiments requiring high accuracy and that are executed on Raspberry Pis. The main channel of the Monsoon power meter needs to be connected to the GPIO pins of the Raspberry Pi, namely a 5V source pin and a pin for the ground (GND). Users should be careful in supplying the correct voltage and avoid current spikes that can damage the Raspberry Pi. A protocol providing step-by-step instructions for using the equipment is crucial to avoid misuse. During our course, it also happened that students accidentally left software running on cluster nodes or used nodes concurrently. We tell students to stop all software that is launched after finishing their session and to check running processes before carrying out their experiment. Concurrent access is mitigated by using a shared calendar in which users book specific time slots on cluster nodes. Software installed on cluster nodes, smartphones, and development boards can also influence experiment execution, if not maintained. Different versions of tools and libraries can cause incompatibilities. Legacy versions of a tool or operating system can be less efficient and contain security flaws, and thus, influence experiment outcomes. Older versions of operating systems may not support newer versions of the software to be tested by users.  
\section{Conclusions}\label{sec:conclusions}

The main contribution of this chapter is the description of the Green Lab course, \ie a course in the area of empirical software engineering in the context of energy-efficient software.
The design, organization, and syllabus of the course can be used as a basis for delivering a similar course in other universities and programs. The provided online material should facilitate a replication of the course with relatively low effort. Our recommendations and lessons learned reported in Section~\ref{sec:lessons_learned} are the results of ten years of experience giving the Green Lab and should help our peers in delivering an even better course than ours; several of those lessons learned may also apply to courses outside the ESE research area.  

As future improvements of the course, we are evaluating the possibility of (i) integrating the usage of the Empirical standard \cite{ralph2020empirical}, both as educational material and as an assessment tool (both for instructors and peer assessment), (ii) anticipating the discussion of threats to validity earlier~\cite{Verdecchia2023} within the timeline of the course (before the deadline for Assignment 2), and (iii) complementing the Green Lab with a more general course about ESE research methods.

Finally, we invite the ESE community to adopt (or even improve) our open-source \textit{tools} for supporting the execution of measurement-based experiments (see Section~\ref{sec:tools}). Over the years, they allowed our research group and our students to save a \textit{considerable} amount of time and to adhere by design to well-known empirical guidelines with very low effort; we are actively maintaining those tools and we are open for feedback, comments, and possible collaborations about them.   

\section*{Acknowledgements}
We would like to thank all students and teaching assistants of the course over the years for having contributed to the development and various improvements of the Green Lab course (and its accompanying tools). We would like to thank Giuseppe Procaccianti for having initiated this amazing and highly rewarding course; and our industrial partners that donated technologies for the course initial setup. We would like to thank Massimiliano Di Penta, Marco Torchiano, and Giuseppe Procaccianti
for having supported the course with original material (\eg slides) we built upon when preparing some components of the course. We thank also all guest speakers who contributed to the success of the course with their guest lectures. Finally, we thank the Computer Science department of the Vrije Universiteit Amsterdam for providing the organizational and financial support to run the course. 

This project is partially supported by (i) the LETSGO Project, promoted by the Netherlands Enterprise Agency (Rijksdienst voor Ondernemend Nederland) and (ii) the European Union’s Horizon 2020 research and innovation programme under the Marie Skłodowska-Curie grant agreement No 871342 ‘‘uDEVOPS’’.

\bibliographystyle{spmpsci}
\bibliography{references}
\end{document}